\documentclass[%
reprint,
superscriptaddress,
amsmath,amssymb,
aps,longbibliography,
]{revtex4-1}

\usepackage{graphicx} %
\usepackage{verbatim}
\usepackage[version=4]{mhchem}
\usepackage{xcolor}
\usepackage{physics}
\usepackage[normalem]{ulem}
\usepackage{pdfpages}
\usepackage{pgffor}
\usepackage{booktabs}
\usepackage{xr}

\definecolor{Grey}{rgb}{0.50,0.50,0.50}
\definecolor{Blue}{rgb}{0.00,0.00,1.00}
\definecolor{Red}{rgb}{1.00,0.00,0.00}
\definecolor{Green}{rgb}{0.20,0.80,0.20}
\definecolor{Magenta}{rgb}{0.60,0.00,0.60}
\definecolor{BluBondi}{rgb}{0.00,0.58,0.71}
\definecolor{Orange}{rgb}{0.95,0.46,0.17}
\definecolor{tangerine}{rgb}{0.944,0.522,0}

\newcommand{\editor}[2]{%
  \expandafter\newcommand\csname #1note\endcsname[1]{%
    \textcolor{#2}{{\it (\textbf{#1:} ##1)}}}%
  \expandafter\newcommand\csname #1\endcsname[1]{%
    \textcolor{#2}{##1}}%
  \expandafter\newcommand\csname #1cancel\endcsname[1]{%
    \textcolor{#2}{\sout{##1}}}%
  \expandafter\newcommand\csname #1change\endcsname[2]{%
    \textcolor{#2}{\sout{##1} ##2}}%
  \newenvironment{#1text}{\color{#2}}{\color{black}}
}

\editor{Sofiia}{Blue}

\makeatletter
\AtBeginDocument{\let\LS@rot\@undefined}
\makeatother

\begin{document}

\title{Hierarchical universal open structure dataset for atomistic modeling}

\author{Sofiia Chorna}
\email{sofiia.chorna@epfl.ch}
\affiliation{Laboratory of Computational Science and Modeling, Institut des Mat\'eriaux, \'Ecole Polytechnique F\'ed\'erale de Lausanne, 1015 Lausanne, Switzerland}

\author{Cesare Malosso}
\affiliation{Laboratory of Computational Science and Modeling, Institut des Mat\'eriaux, \'Ecole Polytechnique F\'ed\'erale de Lausanne, 1015 Lausanne, Switzerland}

\author{Jonathan Schmidt}
\affiliation{Laboratory of Computational Science and Modeling, Institut des Mat\'eriaux, \'Ecole Polytechnique F\'ed\'erale de Lausanne, 1015 Lausanne, Switzerland}

\author{Michele Ceriotti}
\email{michele.ceriotti@epfl.ch}
\affiliation{Laboratory of Computational Science and Modeling, Institut des Mat\'eriaux, \'Ecole Polytechnique F\'ed\'erale de Lausanne, 1015 Lausanne, Switzerland}

\newcommand{\CSM}[1]{{\color{red}#1}}
\newcommand{\rrscan}{r$^{2}$SCAN}

\date{\today}%

\begin{abstract}
Atomic-scale simulations have been transformed by machine-learning models that predict the properties of atomistic structures with the accuracy of quantum calculations, but at a small fraction of the cost. 
These models have increased dramatically in generality, thanks to the availability of synthetic datasets of ever-increasing size and scope.
The proliferation of datasets, with properties often computed using different levels of electronic-structure approximation, is problematic, as there is often overlap between different datasets, and strong correlations between structures obtained with short molecular dynamics trajectories or geometry optimization.
To address these problems, we introduce a protocol to stratify a dataset into a hierarchy of nested subsets, subsampling from the coarsest to the finest resolution based on an ``ecumetric'' built to reflect the latent space representation of several popular universal machine-learning interatomic potentials. 
We then use this approach to combine most of the open structural datasets currently available, consolidating them into MAD-COSMO, a well-organized collection of almost $1.7$ billion structures, whose $2^{24}$ most diverse members form the MAD-CORE stratified coreset, which can be easily used for all sorts of tasks, from exploratory experiments to the training of foundation models at scale.
We envision that this collection will guide and help coordinate future efforts, focusing them on the most relevant and/or undersampled regions of chemical and structural space, and that it will be further extended to incorporate the outcomes of the collective work of this community.

\end{abstract}

\maketitle

\section{Introduction}

Atomistic machine learning potentials have transformed materials simulation by providing near \textit{ab initio} accuracy at a fraction of the computational cost of electronic-structure methods \cite{behl-parr07prl, deri+19am, behl16jcp}. More recently, the emergence of universal machine learning interatomic potentials (uMLIPs) has extended this success across an unprecedented range of chemistries and simulation tasks, fundamentally reshaping the field \cite{chen-ong22ncs,bata+22nips,mazi+25ncomm, Wood2026, Zhang2026, Rhodes2025,Bochkarev2024}. A central driver of this progress has been the availability of large-scale datasets of density functional theory (DFT) calculations, many of which now contain millions of structures across diverse chemistries and application domains \cite{Schmidt2024, Deng2023, Cavignac2025, Tran2023, Kaplan2025, Eastman2023, Barroso2024, Levine2025OMol25, Levine2026OPoly26}. At the same time, the public atomistic data ecosystem has developed in a largely decentralized manner. Consequently, datasets and benchmarks are often aggregated from heterogeneous sources with different electronic-structure settings and convergence criteria, leading to inconsistent labels and substantial numerical noise \cite{Bosoni2024, Prandini2018, Mazdziarz2024, Lam2026, Kuryla2025, Rossomme2023, Lejaeghere2016}. In addition, many datasets contain overlapping or near-identical structures, such that aggregating them can unintentionally over-represent certain regions of configuration space \cite{Siron2025, Martirossyan2025}. Building a coherent and diverse dataset is then not just a matter of concatenation but rather a technically and conceptually non-trivial task \cite{Qi2024, Kulichenko2024, Zou2026, Yu2026}.

In this work, we address these limitations by constructing a hierarchical view of existing atomistic datasets, in which each level provides provable uniform coverage of the data it is built from, at increasing levels of resolution.
The construction relies on an \textit{ecumenical} metric, a distilled structural distance obtained by training a SPACE model \cite{Bigi2026} on the last-layer features of several popular universal interatomic potentials, obtaining a single learned consensus representation. We first construct the hierarchy on a dataset built based on the principle of Massive Atomic Diversity (MAD) \cite{mazi+25sd,Malosso2026}. 
Then we systematically extend it to other datasets using an \textit{accretion} algorithm that tries to preserve the initial ordering at every level, without rebuilding it from scratch, and continues selection as farthest point sampling. We apply this pipeline to a collection of almost $1.7$ billion structures from a number of major public atomistic datasets, which we refer to as MAD-COSMO, and construct its nested coreset of $2^{24}$ structures, MAD-CORE.

This multi-resolution representation can be helpful for several practical applications. The hierarchical ordering of the library can be used to effectively select representative structures for budget-constrained recalculations at different levels of theory, hyperparameter exploration, and architectural benchmarking. It also gives a quantitative measure of dataset completeness by identifying areas where new \textit{ab initio} calculations can most efficiently improve the coverage of the represented configuration space.

\begin{figure*}[htbp]
    \centering
    \includegraphics[width=\linewidth]{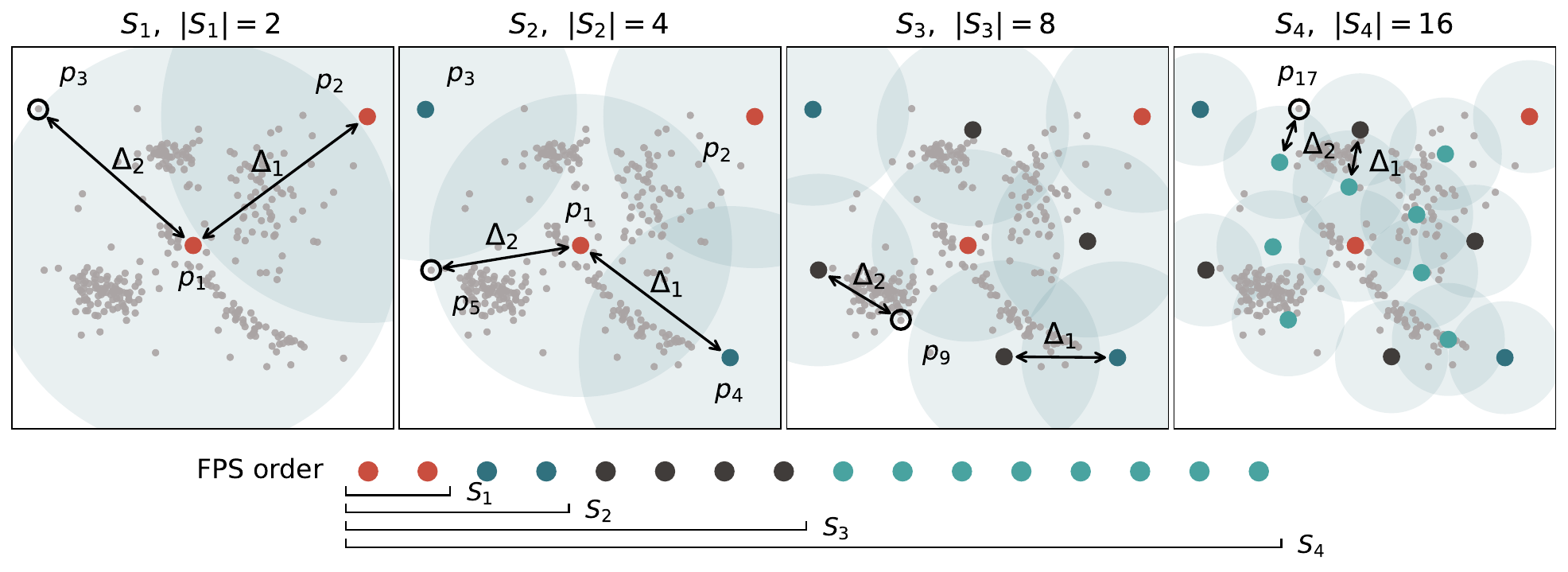}
    \caption{Nested $k$-set hierarchy constructed by FPS on a synthetic 2D point cloud, with $|S_k| = 2^k$. Colors mark the level at which points enter the hierarchy, shaded circles the coverage radius, $\Delta_1$ the separation, and $\Delta_2$ the coverage, i.e., the distance to the next FPS pick (open marker). The lower part of the figure shows the FPS ordering that generates the nested levels $S_1 \subset S_2 \subset S_3 \subset S_4$}
    \label{fig:nested_kset_example}
\end{figure*}

\section{Nested coresets}
\label{sec:nested-coresets}

We begin by introducing an algorithm to build and maintain a ``nested coreset'' structure over a large, heterogeneous dataset. 
The goal is to identify subsets of samples that achieve a progressively higher resolution.
An important requirement is that the framework should include the possibility of adding new structures to an existing corpus, or to re-define the ordering based on a different metric. 
Doing so should not completely change the coarse levels of the hierarchy, as these samples may have already been used to evaluate computationally-demanding property values. 

\subsection{Separation and coverage via farthest point sampling}

We first define the notion of uniform coverage at a given resolution that each level of the hierarchy should provide. Consider a dataset $X$ embedded in the latent space $\mathbb{R}^D$ and a metric $d$ induced by the chosen descriptor. A representative subset $S \subseteq X$ of fixed size can be assessed by its \textit{separation}, the minimum distance between any pair of points in $S$,
\begin{equation}
\mathrm{sep}(S) = \min_{\substack{x,y \in S \\ x \neq y}} d(x, y),
\end{equation}
and by its \textit{coverage} of $X$ by $S$, the maximum distance from any point of $X$ to its nearest neighbor in $S$,
\begin{equation}
\mathrm{cov}_X(S) = \max_{x \in X}\, \min_{s \in S} d(x,s).
\end{equation}

An ideal representative set guarantees that the selected points are mutually well-separated, so each contributes new information, and that they uniformly cover the descriptor space, leaving no region unrepresented. In this context, Farthest Point Sampling (FPS) \cite{gonzalez1985} is a classical greedy algorithm that links separation and coverage by construction. Starting from an arbitrary point $s_1 \in X$, it iteratively samples the point that is the farthest from all current selections:
\begin{equation}
s_{i+1} = \operatorname*{argmax}_{x \in X} \min_{j=1,\dots,i} d(x,s_j).
\end{equation}
As the farthest point is the one that determines the coverage, its distance to the existing set upon inclusion becomes the new minimum distance within the selected set, and therefore the separation of $S$ is never smaller than its coverage of $X$, so
\begin{equation}
\mathrm{sep}(S) \geq \mathrm{cov}_X(S)
\label{eq:separation-coverage}
\end{equation}
holds for every step of the selection.

Minimizing coverage exactly is the NP-hard k-center problem \cite{hsu1979}. Exact solvers now handle up to a million points \cite{Contardo2019, Gaar2022}, but remain far from the scale required in this work. FPS gives a worst-case factor-two approximation to the optimal coverage radius under the triangle inequality \cite{gonzalez1985, hochbaum1985}, meaning that the first $n$ selected points achieve a coverage radius of $X$ no more than twice the minimum possible. Since each selection requires updating the distance to the selected point for all remaining candidates, selecting $n$ points from $N$ candidates requires $\mathcal{O}(N \times n)$ distance evaluations. In practice, this can be accelerated with Voronoi-pruning \cite{cers+21mlst} on CPUs and GPU-parallel distance updates, which makes FPS a popular tool for subset selection in atomistic datasets \cite{mazi+25sd, Trestman2025, Dral2021, De2016}. A key property of FPS that we exploit in this work is that the sequence of points is \textit{nested} by construction, with the first $n$ points being a subset of the first $n+1$, so a single ordering defines a deterministic hierarchy of coresets at every size.

We construct a hierarchical, multi-resolution view of the dataset by retaining power-of-two prefixes of the FPS ordering. A nested $k$-set hierarchy of depth $K$ over $X$ is the sequence: 
\begin{equation}
S_0 \subset S_1 \subset \cdots \subset S_K \subseteq X, \qquad |S_k| = 2^k,
\end{equation}
where $S_0$ holds the single point that seeds the FPS run and each level inherits the separation-coverage guarantee \eqref{eq:separation-coverage} from the underlying FPS. To build intuition, consider an example given in Figure \ref{fig:nested_kset_example} that shows the hierarchy on a synthetic 2D point cloud built via FPS. Each successive level doubles the number of selected points, tightening the coverage.

\begin{figure}[tbp]
    \centering
    \includegraphics[width=\linewidth]{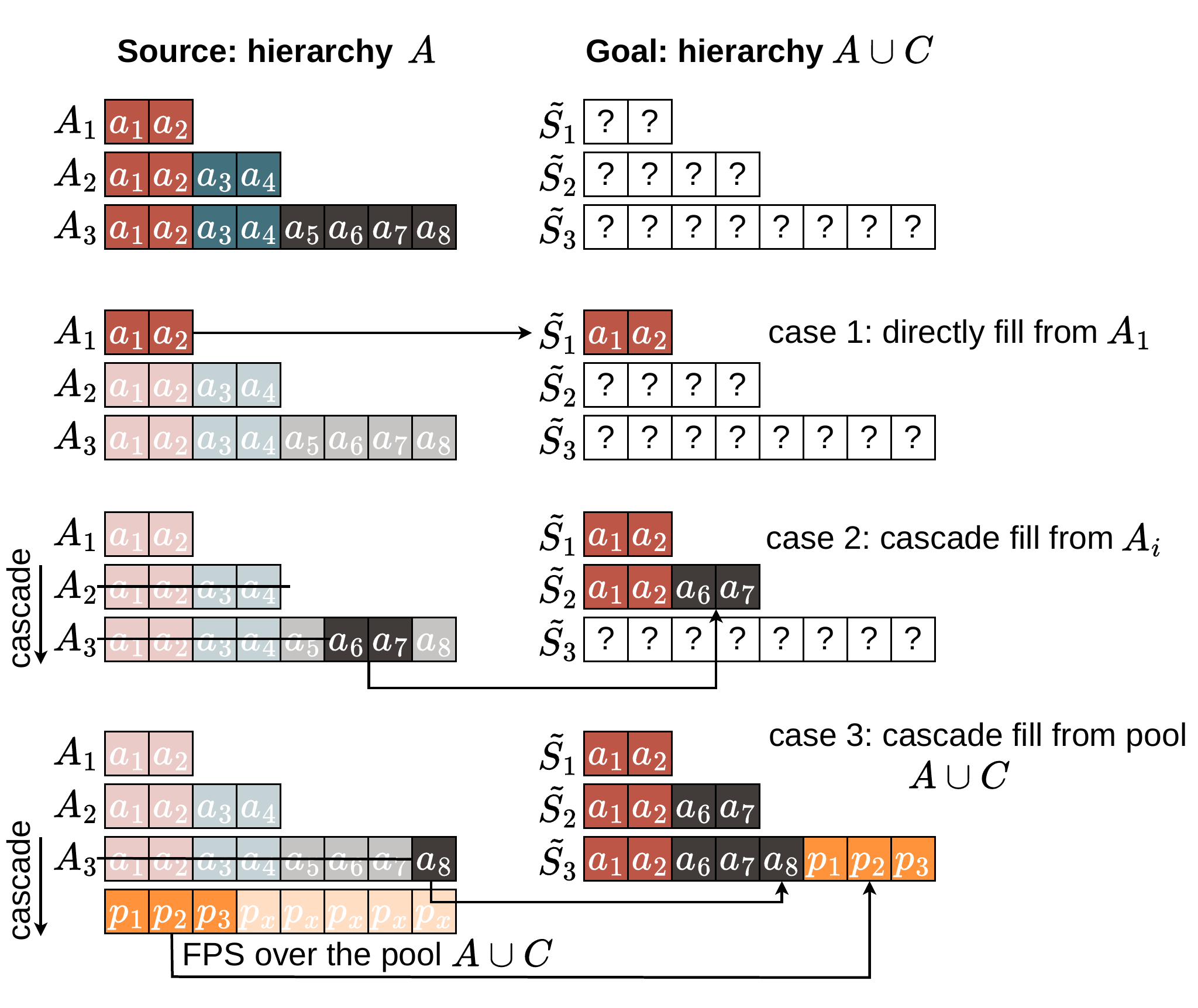}
    \caption{Schematic illustration of the \textit{accretion} algorithm. Each level $\tilde{S}_k$ of the updated hierarchy over $A \cup C$ attempts to keep as many points of the initial hierarchy $A$ as the separation-coverage guarantee \eqref{eq:separation-coverage} allows, drawing first from the corresponding level $A_k$ (case 1), then from deeper levels of the initial hierarchy (case 2), and finally filling the remaining slots by FPS over the combined pool $A \cup C$  (case 3)}
    \label{fig:accrete}
\end{figure}

\subsection{Accretion algorithm}
\label{sec:accrete}

A single FPS run over a collection of datasets would fill the coarse levels with its most extreme structures regardless of their origin. We envisage these levels to be the most informative, and want them to be reusable even if the overall corpus $X$ is further extended. We therefore build the hierarchy on one curated dataset first and then merge in the rest of the collection (Section \ref{sec:building}). We use an \textit{accretion} operation that starts from the initial ordering $A$ and merges a candidate set $C$ into a hierarchy over $A \cup C$, keeping at every level as many initial points as the search finds compatible with guarantee \eqref{eq:separation-coverage}. If every new structure lies within the coverage radius of structures from the initial ordering, the hierarchy will keep its initial members.

The new ordering is built level by level over the combined pool. To grow the selection from $2^{k-1}$ to $2^k$ points, we first re-insert as many points as possible from the initial ordering, starting with its level $A_k$ (the first $2^k$ points of $A$) and cascading down to deeper levels when needed, and let plain FPS over $A \cup C$ fill the remaining slots. The re-inserted points are added farthest-first with respect to the current selection, so \textit{accretion} preserves the membership of the initial hierarchy levels rather than the exact order within them. Keeping more initial points lowers the separation of the level, and at some point it falls below the coverage of the pool and the guarantee \eqref{eq:separation-coverage} breaks. 
We use binary search as a heuristic on how many initial points to keep. One trial keeps a given number of them, completes the level with plain FPS, and checks the guarantee at the end, and its outcome halves the search range, toward larger counts if the guarantee holds and smaller ones otherwise. The accepted count is the largest that passed the check among the trials. Each trial costs at most as much as completing the level once, and the binary search requires only a logarithmic number of trials. 
Initial points that lose their slot at a coarse level stay in the pool and can re-enter at finer levels. The workflow is shown in Figure \ref{fig:accrete}. The resulting ordering $\tilde{S}_k$ is nested, and the guarantee \eqref{eq:separation-coverage} is explicitly verified on every completed level (details are in SI \ref{app:fps-accrete-impl}). 
After accretion, the separation-coverage condition \eqref{eq:separation-coverage} is only fulfilled at the power-of-two levels and not at every intermediate size, which would require a fresh FPS run. 

Beyond organizing the selected points into the hierarchy described previously, all other data can be arranged relative to the coreset by assigning the rest of $X$ to its nearest coreset point. Every selected structure of the coreset then forms a cell of all the structures that lie closer to it than to other selections, namely it creates a Voronoi tessellation. The tessellation exists at every level $S_k$, and the notion of resolution can be further extended to the local level by ordering cell points with FPS as well.

\subsection{Using nested coresets}

Having a hierarchical representation of the coreset and its further tessellation of the whole collection can facilitate many practical tasks. For the universal training set construction, walking through the ordering from the top down to a fixed budget selects the most diverse subset of that size.
We will demonstrate later that this approach allows one to train models of near-uniform accuracy over the full range covered by the global corpus of samples, which is a desirable feature when striving for universality and broad transferability.
If uniform coverage puts too much weight on the sparse extremes, the coreset points can instead be sampled with a probability proportional to the size of their Voronoi cells raised to an exponent between zero and one, which interpolates between uniform coverage of cell centers and the density of the full collection \cite{ceri+13jctc}. 
The sampling can be further filtered or reweighted, for example, by structure size or composition.

Another useful application is to gather training structures for more focused learning tasks. 
Given a few structures corresponding to the problem at hand, one can quickly extract all samples within a prescribed ecumetric distance, which can be done efficiently by finding the coreset points nearest to the target structure and collecting the content of their Voronoi cells. 
A further FPS subselection can be performed if needed. The tessellation also shows whether a structure is already well represented in the collection, as it is enough to inspect the cell of its nearest coreset point instead of searching the whole library.

Last but not least, we envisage that MAD-COSMO and MAD-CORE should be living libraries that can be updated yearly as more open datasets become available and that the community can use to direct efforts to compute high-quality energy, forces and quantum-mechanical properties to train universal models that cover all use cases that have been released as open, FAIR datasets~\cite{roadmap}. 

\section{Building MAD-COSMO}

\subsection{Atomistic datasets}
\label{sec:building}

We begin by consolidating a large fraction of the publicly available atomistic data into a single dataset, which we name MAD-COSMO. 
We collect datasets from a variety of sources that fall under licenses that allow further redistribution and reuse for both academic and commercial use, such as CC BY 4.0, and we exclude the datasets with undefined licenses, or that have been superseded by later releases. 
We deliberately strip all quantum property information and retain only the structures, both because it is essentially impossible to ensure consistency between values computed with different settings, and because we do not wish to replace the original data records, which should still be referenced and used as primary sources by anyone using them for training.
This work would not have been possible at this scale without recent advances in FAIR data infrastructure~\cite{tarl+20sd, Scheidgen2023, Vita2023, Evans2024}. Despite this progress, missing or ambiguous license information remained a barrier preventing the reuse of in principle findable and interoperable data. The collection comprises more than 1000 references~\cite{0, 1, 2, 3, 4, 5, 6, 9, 10, 11, 12, 14, 15, 16,
19, 20, 21, 22, 23, 24, 25, 26, 27, 28, 29, 30, 31, 32, 33, 34, 35, 36, 37, 38, 39, 40,
41, 42, 43, 44, 45, 46, 47, 48, 49, 50, 51, 52, 53, 54, 55, 56, 59, 60, 61, 62,
63, 64, 66, 68, 69, 70, 71, 72, 73, 76, 77, 78, 80, 81, 82, 83, 84, 85, 86,
92, 95, 96, 97, 98, 99, 100, 101, 105, 106, 107,
109, 110, 111, 112, 114, 115, 116, 117, 118, 119, 120, 121, 122, 123, 124, 125,
126, 127, 128, 129, 130, 131, 132, 133, 134, 136, 137, 138, 139, 140, 141, 142,
143, 144, 145, 146, 147, 148, 149, 150, 151, 152, 154, 155, 157, 158, 159,
160, 161, 162, 163, 164, 165, 166, 167, 168, 169, 170, 171, 172, 174, 175, 176,
177, 178, 179, 180, 181, 183, 184, 185, 186, 187, 188, 189, 190, 191, 192, 193,
194, 195, 196, 197, 198, 199, 200, 201, 202, 203, 204, 205, 206, 207, 208, 209, 210,
211, 212, 213, 214, 215, 216, 217, 218, 219, 220, 221, 222, 223, 224, 225, 226, 227,
228, 229, 230, 231, 232, 233, 234, 235, 236, 237, 238, 239, 240, 241, 243, 244, 245,
246, 247, 248, 249, 250, 251, 252, 253, 254, 255, 256, 257, 258, 259, 260, 261, 262,
263, 264, 265, 266, 267, 268, 269, 270, 271, 272, 273, 274, 276, 277, 278, 279, 280,
281, 282, 283, 284, 285, 287, 288, 289, 290, 291, 292, 293, 294, 295, 296, 298,
299, 300, 301, 302, 303, 304, 305, 306, 307, 308, 309, 310, 311, 312, 313, 314, 315,
316, 317, 319, 320, 321, 322, 323, 324, 325, 327, 328, 329, 330, 331, 332, 333, 334,
335, 336, 337, 338, 339, 340, 341, 342, 343, 344, 345, 346, 347, 348, 349, 350, 351,
352, 353, 358, 359, 360, 361, 362, 363, 364, 365, 366, 367, 368,
369, 370, 371, 372, 373, 374, 375, 376, 377, 378, 379, 380, 381, 382, 383, 384, 385,
386, 387, 388, 390, 391, 392, 393, 394, 395, 396, 397, 398, 402, 403, 404, 405, 406,
407, 408, 409, 410, 413, 414, 415, 416, 417, 418, 419, 420, 421, 422, 423, 424, 425,
426, 427, 428, 429, 430, 431, 432, 433, 434, 435, 436, 437, 438, 439, 440, 441, 442,
443, 444, 445, 446, 447, 448, 449, 450, 453, 454, 455, 456, 457, 458, 459, 460,
462, 463, 464, 465, 466, 467, 468, 469, 470, 471, 472, 475, 476, 477, 478, 479, 480,
481, 483, 485, 486, 487, 488, 489, 490, 493, 494, 495, 496, 497, 500, 501, 502,
503, 504, 505, 506, 507, 508, 509, 510, 511, 512, 514, 515, 516, 517, 518, 519, 520,
522, 524, 525, 526, 527, 528, 529, 530, 531, 533, 535, 536, 537, 539, 541, 543, 544,
545, 546, 547, 548, 549, 551, 552, 553, 555, 556, 557, 558, 559, 561, 562, 563, 564,
565, 567, 568, 569, 570, 571, 572, 573, 574, 575, 576, 577, 578, 579, 580, 582, 583,
584, 585, 586, 587, 588, 590, 591, 592, 593, 594, 596, 597, 598, 599, 600, 601, 602, Johnson2022CCCBDB, Gupta2021, Grisafi2022, jain+13aplm, cf_0, cf_1, cf_2, cf_3, cf_4,
cf_5, cf_6, cf_7, cf_8, cf_9, cf_10, cf_11, cf_12, cf_13, cf_14, cf_15, cf_16, cf_17,
cf_18, cf_19, cf_20, cf_21, cf_22, cf_23, cf_24, cf_25, cf_26, cf_27, cf_28, cf_29,
cf_30, cf_31, cf_32, cf_33, cf_34, cf_35, cf_36, cf_37, cf_38, cf_39, cf_40, cf_41,
cf_42, cf_43, cf_44, cf_45, cf_46, cf_47, cf_48, cf_49, cf_50, cf_51, cf_52, cf_53,
cf_54, cf_55, cf_56, cf_58, cf_59, cf_60, cf_61, cf_62, cf_63, cf_64, cf_65, cf_66,
cf_67, cf_68, cf_69, cf_70, cf_71, cf_72, cf_73, cf_74, cf_75, cf_76, cf_77, cf_78,
cf_79, cf_80, cf_81, cf_82, cf_83, cf_84, cf_85, cf_86, cf_87, cf_89, cf_90, cf_91,
cf_92, cf_93, cf_94, cf_95, cf_96, cf_97, cf_98, cf_99, cf_100, cf_101, cf_102, cf_103,
cf_104, cf_105, cf_106, cf_107, cf_108, cf_109, cf_110, cf_111, cf_112, cf_113, cf_114,
cf_115, cf_116, cf_117, cf_118, cf_119, cf_120, cf_121, cf_122, cf_124, cf_125, cf_126,
cf_127, cf_128, cf_129, cf_130, cf_131, cf_132, cf_133, cf_134, cf_135, cf_136, cf_137,
cf_138, cf_139, cf_140, cf_142, cf_143, cf_144, cf_145, cf_146, cf_147, cf_148, cf_149,
cf_150, cf_151, cf_152, cf_153, cf_154, cf_155, cf_156, cf_157, cf_158, cf_159, cf_160,
cf_161, cf_162, cf_163, cf_164, cf_165, cf_166, cf_167, cf_168, cf_169, cf_170, cf_171,
cf_172, cf_173, cf_174, cf_175, cf_176, cf_177, cf_178, cf_179, cf_180, cf_181, cf_182,
cf_183, cf_184, cf_185, cf_186, cf_188, cf_189, cf_190, cf_191, cf_192, cf_193, cf_194,
cf_195, cf_196, cf_197, cf_198, cf_200, cf_201, cf_202, cf_203, cf_204, cf_205, cf_206,
cf_207, cf_208, cf_209, cf_210, cf_211, cf_213, cf_214, cf_215, cf_216, cf_217, cf_219,
cf_220, cf_221, cf_222, cf_223, cf_224, cf_225, cf_226, cf_227, cf_228, cf_229, cf_230,
cf_231, cf_233, cf_234, cf_235, cf_236, cf_237, cf_238, cf_239, cf_240, cf_241, cf_242,
cf_243, cf_244, cf_245, cf_246, cf_247, cf_248, cf_249, cf_250, cf_251, cf_252, cf_253,
cf_254, cf_255, cf_256, cf_257, cf_258, cf_259, cf_260, cf_261, cf_262, cf_263, cf_264,
cf_265, cf_266, cf_267, cf_268, cf_269, cf_270, cf_271, cf_273, cf_274, cf_275, cf_276,
cf_277, cf_278, cf_279, cf_280, cf_281, cf_282, cf_283, cf_284, cf_285, cf_286, cf_287,
cf_288, cf_289, cf_290, cf_291, cf_292, cf_293, cf_294, cf_295, cf_296, cf_297, cf_298,
cf_299, cf_300, cf_301, cf_302, cf_303, cf_304, cf_306, cf_307, cf_308, cf_309, cf_310,
cf_311, cf_312, cf_313, cf_314, cf_315, cf_316, cf_317, cf_318, cf_319, cf_320, cf_321,
cf_322, cf_323, cf_324, cf_325, cf_326, cf_327, cf_328, cf_329, cf_330, cf_331, cf_332,
cf_333, cf_334, cf_335, cf_336, cf_337, cf_338, cf_339, cf_340, cf_341, cf_342, cf_343,
cf_344, cf_345, cf_346, cf_347, cf_348, cf_349, cf_350, cf_351, cf_352, cf_353, cf_354,
cf_355, cf_356, cf_357, cf_358, cf_359, cf_360, cf_361, cf_362, cf_363, cf_364, cf_365,
cf_366, cf_367, cf_368, cf_369, cf_370, cf_371, cf_372, cf_374, cf_375, cf_376, cf_377,
cf_378, cf_379, cf_380, cf_381, cf_382, cf_383, cf_384, cf_385, cf_386, cf_387, cf_388,
cf_389, cf_390, cf_391, cf_393, cf_394, cf_395, cf_396, cf_397, cf_398, cf_399, cf_400,
cf_401, cf_402, cf_403, cf_404, cf_405, cf_406, cf_407, cf_408, cf_409, cf_410, cf_411,
cf_412, cf_413, cf_414, cf_415, mcloud_00s2s_xre47, mcloud_02rcd_nc341,
mcloud_05j11_ky482, mcloud_06329_0n866, mcloud_07wfy_shs16, mcloud_08a12_p4a37,
mcloud_0db0e_t6d44, mcloud_0ewdw_fx861, mcloud_0p5ya_w1n45, mcloud_16jd7_0zz18,
mcloud_1ap7q_cn115, mcloud_1bymd_by722, mcloud_1fx58_pde31, mcloud_1herw_7m923,
mcloud_1qfe8_htw65, mcloud_1z76q_e6n84, mcloud_1z_pd, mcloud_1zpzw_a7p58,
mcloud_238ad_1bg44, mcloud_286v9_64865, mcloud_2aj5h_tzk43, mcloud_2cn5m_gek76,
mcloud_2daf9_7pk21, mcloud_2e2md_b4070, mcloud_2ftdw_8cr18, mcloud_2gr2g_h8s29,
mcloud_2kh6a_57d12, mcloud_2y89r_m1q15, mcloud_342fz_5wx57, mcloud_36x8p_hrv47,
mcloud_3a62g_a7r35, mcloud_3agdy_69a84, mcloud_3t20t_26564, mcloud_3ydnb_ckz62,
mcloud_3yf97_hzb15, mcloud_3yke8_xxf82, mcloud_405an_d8183, mcloud_45yfz_7xa87,
mcloud_4m62e_np961, mcloud_4t64a_r3d56, mcloud_4vt5h_sjd96, mcloud_55jga_d0447,
mcloud_57kc3_fhs52, mcloud_58a61_7te73, mcloud_5jsfj_cdg70, mcloud_5pdz3_vsq42,
mcloud_5vh19_61y09, mcloud_61kg0_hy694, mcloud_645qm_6ss59, mcloud_6eedj_3ey43,
mcloud_6f23w_zqe90, mcloud_6jc67_ars95, mcloud_6ntbp_vbx54, mcloud_6t3hd_9v682,
mcloud_6t5yd_kek31, mcloud_6txnj_70460, mcloud_6xm49_rz892, mcloud_71861_xfe58,
mcloud_7p3z3_t3q91, mcloud_7qm51_anq57, mcloud_7v6p9_92t31, mcloud_7v74k_sbq73,
mcloud_83e5g_ts228, mcloud_8a0qm_ae178, mcloud_8fdar_jpw03, mcloud_8fj15_e3949,
mcloud_8q9cy_0e333, mcloud_8qvdy_t3662, mcloud_8sbz4_6ng07, mcloud_8z1k0_m0p53,
mcloud_935a7_wb166, mcloud_95z7z_zbr43, mcloud_9hydd_z0v77, mcloud_9j1pn_77r98,
mcloud_9qf6r_9fq18, mcloud_9x8ac_2dg09, mcloud_a68hf_tt080, mcloud_a8r4c_whw21,
mcloud_afx8g_ra033, mcloud_ann3c_0qq90, mcloud_anq8a_2ep59, mcloud_as20t_exy47,
mcloud_as73a_ja910, mcloud_at71v_gdh04, mcloud_b0ved_e1358, mcloud_b4kgx_rtd78,
mcloud_b9k22_a4856, mcloud_bj15t_p3415, mcloud_bj9ac_bn156, mcloud_bmrfs_z1892,
mcloud_bnxsn_7w947, mcloud_bzch9_1xh79, mcloud_c00et_26a37, mcloud_c0hpy_pad70,
mcloud_c3px7_y0k21, mcloud_c40dt_dg536, mcloud_c6ngz_zrh49, mcloud_c9ksd_mbb05,
mcloud_c9peh_kt938, mcloud_cdh59_rfp34, mcloud_cmh66_t8h09, mcloud_cr6yv_kmm28,
mcloud_crxpj_3gk87, mcloud_d6pmr_tty23, mcloud_d7k4k_xbw11, mcloud_dby42_mrn61,
mcloud_dca0t_q7m31, mcloud_desfh_3ma25, mcloud_detxn_0ga89, mcloud_dfnmr_0r807,
mcloud_dr2cc_f2047, mcloud_drzx4_bh631, mcloud_dyvt8_af249, mcloud_e4fat_6ga07,
mcloud_ed_g2, mcloud_eg3pn_1fw83, mcloud_emryk_yfa58, mcloud_es05d_9jc34,
mcloud_et7sa_3cv85, mcloud_f20vs_e0v74, mcloud_f4cjq_y4987, mcloud_f59hy_ben23,
mcloud_ffkqx_5ae17, mcloud_ffvcy_q4c20, mcloud_fmrcj_08z10, mcloud_fx32t_12t34,
mcloud_g5rqp_2aw69, mcloud_g8jhr_t8x36, mcloud_grmx8_xva78, mcloud_grpwm_0mn79,
mcloud_gwxqt_erf27, mcloud_hacw9_yzc96, mcloud_hbpvk_pfp22, mcloud_hdb97_me114,
mcloud_hk0x5_xxq93, mcloud_hmdsb_k3h43, mcloud_hnjfh_nd118, mcloud_hqa6g_z2386,
mcloud_hry14_dk842, mcloud_hv7b1_71340, mcloud_hx49x_ab323, mcloud_hyjhn_8em96,
mcloud_j144f_74t50, mcloud_j2fka_sda13, mcloud_j2myj_5zd34, mcloud_j4v22_z2c33,
mcloud_jk_9v, mcloud_jsrvx_8yp49, mcloud_jwt9m_dtt37, mcloud_jxf1h_pe887,
mcloud_jyjyw_fm563, mcloud_k1hjd_y6e90, mcloud_k45dq_ebf35, mcloud_k79n4_gaj04,
mcloud_kacdj_ssv28, mcloud_kbsns_73a82, mcloud_kd6ct_85z54, mcloud_kf8d0_27289,
mcloud_kqv0y_v9k76, mcloud_kwfgp_94v27, mcloud_kyaqv_avg31, mcloud_m0_zg,
mcloud_m17qc_y7350, mcloud_m7d8t_ktv16, mcloud_m9gfa_xr118, mcloud_n271h_74987,
mcloud_nc4x8_yme87, mcloud_nepw0_kn826, mcloud_nf76v_1eh14, mcloud_njdbv_4yr45,
mcloud_nqd1f_q1z28, mcloud_nr4an_byf21, mcloud_nsp7v_2dk91, mcloud_nz8sk_6h010,
mcloud_p8n7a_p5e75, mcloud_pke3c_65m49, mcloud_pkhy1_ted45, mcloud_pkn2d_h8190,
mcloud_pp9m5_dfr36, mcloud_pqa4c_qk811, mcloud_q2smg_en715, mcloud_qbdsr_k2m78,
mcloud_qe7x5_fbb10, mcloud_qffdm_h9w98, mcloud_qp1ej_vst22, mcloud_qs0kz_6t805,
mcloud_qwg2b_ncm42, mcloud_qyb0c_q8m61, mcloud_r0ysk_tx858, mcloud_r8e44_wf170,
mcloud_rnve6_eas10, mcloud_rs6a8_47972, mcloud_s3nkw_qv745, mcloud_s48qh_yzy40,
mcloud_sbvxf_73e81, mcloud_skbnr_9wx61, mcloud_sqrdx_ebj42, mcloud_sqstg_vcd29,
mcloud_t5nvc_rah21, mcloud_tbphj_bsk15, mcloud_tcrks_ymp88, mcloud_txxzk_5mp05,
mcloud_vbd5n_pnp32, mcloud_vermx_08476, mcloud_vpyk8_xaa86, mcloud_vtahk_cd853,
mcloud_vv6sy_98m49, mcloud_w7hen_cme26, mcloud_wpz35_q2y88, mcloud_wwrkx_77334,
mcloud_wxy8k_14m62, mcloud_x1n3f_vwz18, mcloud_xavbv_5h791, mcloud_xb7gc_shq49,
mcloud_xc3bm_qc060, mcloud_xfsv2_2fq31, mcloud_xrtkn_tzm42, mcloud_y84fb_t4h71,
mcloud_y90bb_bpd55, mcloud_yayv3_eha98, mcloud_ydn09_ngs56, mcloud_ykwg5_vb523,
mcloud_yzhmw_vje16, mcloud_z3g78_sq788, mcloud_z6g6d_mp618, mcloud_zd984_2ja12,
mcloud_zhab2_1tp79, mcloud_zne6q_a4h48, mcloud_zqxj1_45w58, mcloud_zt_z4,
mcloud_ztd8f_vj464, mcloud_zw0nd_wx180, mcloud_2dtopo, mcloud_autowannier,
mcloud_curated_cofs, mcloud_pyrene_mofs, mcloud_stoceriaitf,
mcloud_tin_antimony_sulfoiodide, mcloud_mc2d, Saal2013_OQMD,
Kirklin2015_OQMD}, gathered from the ColabFit repository~\cite{Vita2023}, the OPTIMADE API~\cite{Evans2024}, NOMAD~\cite{Scheidgen2023}, Materials Cloud~\cite{mcloud_talirz2020} and complemented by several standalone releases, among them Alexandria~\cite{Cavignac2025, Loew2025, Wang2023}, OMat24~\cite{Barroso2024}, OMol25~\cite{Levine2025OMol25}, the Open Catalyst families OC20~\cite{Chanussot2021}, OC22~\cite{oc22_dataset}, OC25~\cite{oc25}, OC20-NEB~\cite{oc20-neb}, ODAC25~\cite{sriram2025odac25}, and GDB-13~\cite{Blum2009}. The only exception to the licensing rule is GDB-13 published as SMILES strings, which we converted to three-dimensional structures and included with the permission of its authors. The details of the dataset retrieval and preprocessing are given in SI \ref{app:data-collection}. We organize those individual releases into 454 groups, so that datasets that belong to one family, for example MatPES-PBE and MatPES-\rrscan{} or all subsets of Alexandria, as well as all splits of the same dataset, are treated as one dataset group.
The accumulated datasets likely contain duplicated structures of common molecules and materials as different sources recalculate the same systems.
It also contains a large number of molecular structures from GDB-13 that cover a comparatively narrow composition range, which we deliberately included to assess the behavior of the selection protocol when using a corpus with widely varying sampling densities.
Duplicates that exist across different uploads are harmless to the construction of the ordering, as the FPS procedure brings them to the end of the ordering once one copy has been selected, since they have zero distance to the selection set. Together with the MAD-NEXT dataset (Section \ref{sec:ecumenical-metric}), we refer to this collection of 1,694,434,380 structures as MAD-COSMO.

Even though we release MAD-COSMO in an unfiltered form, we find that it contains a few configurations with geometries that are too extreme to be of use in any foreseeable application, and that are likely the result of clerical errors. 
We tag these structures and exclude them from the pool of configurations used to construct MAD-CORE, as otherwise they would dominate the coarsest levels in view of their extreme distortions. 
We identify unphysical geometries using a Ziegler–Biersack–Littmark (ZBL) filter \cite{Ziegler1985}, which is an inexpensive analytical potential that models short-range interactions between nuclei and can be used to detect pathological structures with atoms sitting too close to each other. 
A structure is excluded if its maximum ZBL force is at least $200$ eV/\AA, its ZBL energy per atom exceeds $50$ eV/atom or its maximum atomic ZBL energy is greater than $100$ eV/atom. This filtering excludes extremely compressed structures that would otherwise appear at the top of the ordering, in total less than $0.02\%$ of the collection.
Keeping the filtering rather minimal and excluding only pathological configurations serves the purpose of letting the unusual structures pass. We additionally exclude from the selection a collection of volume scans from NOMAD \footnote{Spread over 21 uploads from 2016 totaling 289,934 entries of equation-of-state scans of binary compounds, with over a thousand configurations per composition. No dataset DOI is associated with these entries. The NOMAD upload identifiers are \texttt{pfQbFCpuRZ-VsVW\_IJV7lQ}, \texttt{ZjmeLpGYRgOgRHgCQAflKQ}, \texttt{iF5Y7WouTRGd68KfVHJzbA}, \texttt{qSCiRCcYSaifBVn00EFOQA}, \texttt{QtM07aJvSGy9dn1OfX5-sA}, \texttt{0DOVoEFyRnaVHvNE0w25yA}, \texttt{m50lbvjYS\_aHm\_aqBmyLGQ}, \texttt{w8VbMKDwSb27iSM2oCE4Kg}, \texttt{nXIM-kNGSNOOBFY7swQlPA}, \texttt{wiBBIXLPRjyQodVo-RALUQ}, \texttt{hP2Col-kS\_WQsMSSPKGUtA}, \texttt{Ihxb9bX4RMSsaPhK017pyA}, \texttt{G\_MmFHcRR6u1niMS3Ky5\_Q}, \texttt{DAF0X3ktS86h1A80\_WFwJg}, \texttt{CyMGpjt6RwmEvfhsn4ZFiA}, \texttt{XuQgFPI0SAixWW5R1Khzqw}, \texttt{RgB-Fx6hQymSwhR36h9ShQ}, \texttt{TRaN13epT4WCn2xT3PY-IQ}, \texttt{5yF9BaWaQ8mB3uEaUevPvQ}, \texttt{8W0cWziPT2ada5XFHSIdjA} and \texttt{0e8exE9ZQuGmo2TOpJVqHw}, whose entries can be retrieved via \texttt{nomad-lab.eu/prod/v1/api/v1/entries?upload\_id=<upload-id>}.} as they pass the filter for highly compressed structures, but would flood the early selections, lying far from other structures even though they amount to large volume distortions of the same systems. All flagged structures remain in MAD-COSMO but are excluded from the selection. Structures with more than 1000 atoms were not featurized and are not part of the collection.
We believe that a similar, conservative level of preliminary data curation will be necessary when updating MAD-COSMO with new structures.
The present snapshot is largely built of structures that were used or designed with a practical application in mind, and therefore reflects a notion of ``usefulness'', and attempts at active structure generation should be designed carefully to avoid flooding the record with structures that maximize diversity in an adversarial way, without any consideration of the potential use. 

\begin{figure}[htbp]
    \centering
    \includegraphics[width=\linewidth]{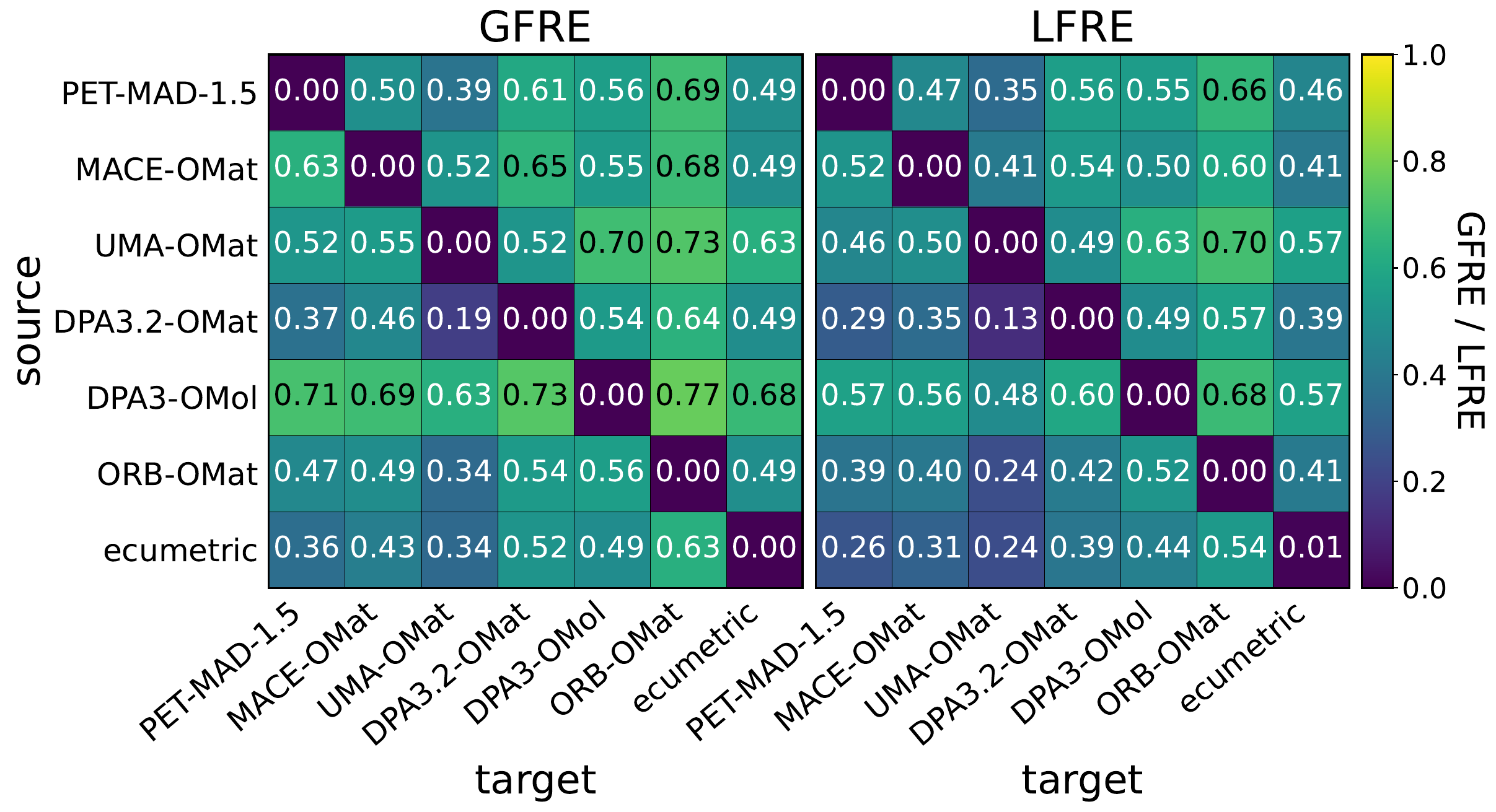}
    \caption{GFRE (left) and LFRE (right) for six uMLIPs and \textit{ecumetric} computed on the structure descriptors over the MAD-NEXT test split. The entry in row $i$ and column $j$ is the error of reconstructing the target features $j$ from the source features $i$. Lower values mean better reconstruction}
    \label{fig:gfre_lfre}
\end{figure}

\subsection{Defining an \textit{ecumetric}}
\label{sec:ecumenical-metric}

The FPS ordering is conditional on the underlying metric used to represent data points. Different structural descriptors or learned embeddings encode a different notion of similarity, imposed by differences in model architecture, training data, and targets \cite{Chorna2026}. 
The same dataset can then result in different hierarchies depending on the representation, as there is no unique model-independent FPS ordering. Since we aim to cover as many atomistic datasets as possible, whose chemical domains are unevenly represented by existing models, choosing a single potential as the latent space might bias the metric to yield a selection that is less effective for other models. To address this source of bias, we instead build an ``ecumetric'' (an \textit{ecumenical metric}) in the spirit of a ``universal'' descriptor.

We train a SPACE model \cite{Bigi2026}, an equivariant message-passing neural network, on the \rrscan{} energies and forces of MAD-NEXT, a dataset composed of MAD-1.5 \cite{Malosso2026} and a collection of new configurations currently being used to prepare MAD-2~\cite{mad-next}, alongside the last-layer atomic features of five uMLIPs taken on these datasets as additional regression targets. We use an equivariant architecture and take the invariant ($L=0$) channels of its backbone as latent features, so that rotated copies of the same structure map to the exact same point of the metric space. The energy and forces link the representations to the potential energy surface, and the auxiliary feature heads regress toward the embedding space of the teacher models. The target feature spaces were chosen to represent the large public datasets and  different levels of electronic structure theory, from GGA \cite{perd+96prl} (MACE-OMat \cite{bata+22nips}, UMA-OMat \cite{Wood2026}, DPA3.2-OMat \cite{Zhang2026}), through meta-GGA \cite{Sun2015} (PET-MAD-1.5 \cite{Malosso2026}) to hybrid functionals \cite{Mardirossian2016} (DPA3-OMol \cite{Zhang2026, DPA3OMol}). For systems that fall outside of a given model's domain, its target features are set to the batch mean during training, and all targets are fitted with a Huber loss to reduce the impact of outliers. Having a single \textit{ecumetric} also serves a practical purpose, as the entire collection of about $1.7 \times 10^9$ structures (Section \ref{sec:building}) can be featurized with one forward pass per structure whereas constructing the common representation from post-hoc alignment of the latent spaces \cite{Moschella2023, Li2026} would require evaluating each structure with multiple models, which is impractical at this scale and complicated to extend for each new dataset. For each structure, we pool its per-atom SPACE backbone features into a structure descriptor by concatenating their mean and standard deviation over the atoms. The standard deviation captures the heterogeneity of the atomic environments within a structure, which mean pooling alone would miss \cite{Chorna2026}. 
This pooling emphasizes the diversity of structures containing few atoms or environments, which is beneficial as it prioritizes small structures, for which calculations can be performed at lower computational effort.
Higher moments or multiple cluster representatives \cite{Iwasaki2026} could retain more environment details but would increase the dimension of the representation which should remain compact for practical use at this scale. We apply a global scaling fitted on the mean and standard deviation of MAD-NEXT to place all structures in a common metric space, in which we use the Euclidean distance. This scaling would not change the FPS picks as it is not applied per-component but globally. Architecture and training protocol are detailed in SI \ref{app:SPACE-training}.

As the main validation criterion, we use the global (GFRE) and local (LFRE) feature reconstruction error metrics of Refs.~\cite{gosc+21mlst,Chorna2026}, which measure how well one feature space (target) can be linearly approximated from another (source). We evaluate all model pairs on the held-out MAD-NEXT test split, pooling per-atom features of every model into a concatenation of mean and standard deviation structure descriptors (Figure \ref{fig:gfre_lfre}). Despite its very limited training effort in comparison to all other universal models, the \textit{ecumetric} is the best source for reconstructing every other feature space under both metrics, with the single exception of UMA-OMat, best reconstructed by DPA3.2-OMat.
We also include a model whose embeddings were not used as regression targets in the training of SPACE, ORB-OMat (orb-v3-conservative-inf-omat) \cite{Rhodes2025}, to have an intuition of extrapolation to unseen feature spaces. Here, SPACE reconstructs the ORB-OMat features with errors comparable to or lower than those of the models actually trained on the same OMat24 dataset.
Even though the \textit{ecumetric} model is not meant to compete with state-of-the-art uMLIPs, its predicted \rrscan{}  energies and forces are reasonably accurate (16 meV/atom energy and 109 meV/\AA{} force MAE on a test set of about 1000 randomly selected MAD-COSMO structures) and can be used as labels for pre-training exercises \cite{Gardner2024, Matin2025, Gardner2025}.

\section{Building MAD-CORE}

\begin{figure}[htbp]
    \centering
    \includegraphics[width=\linewidth]{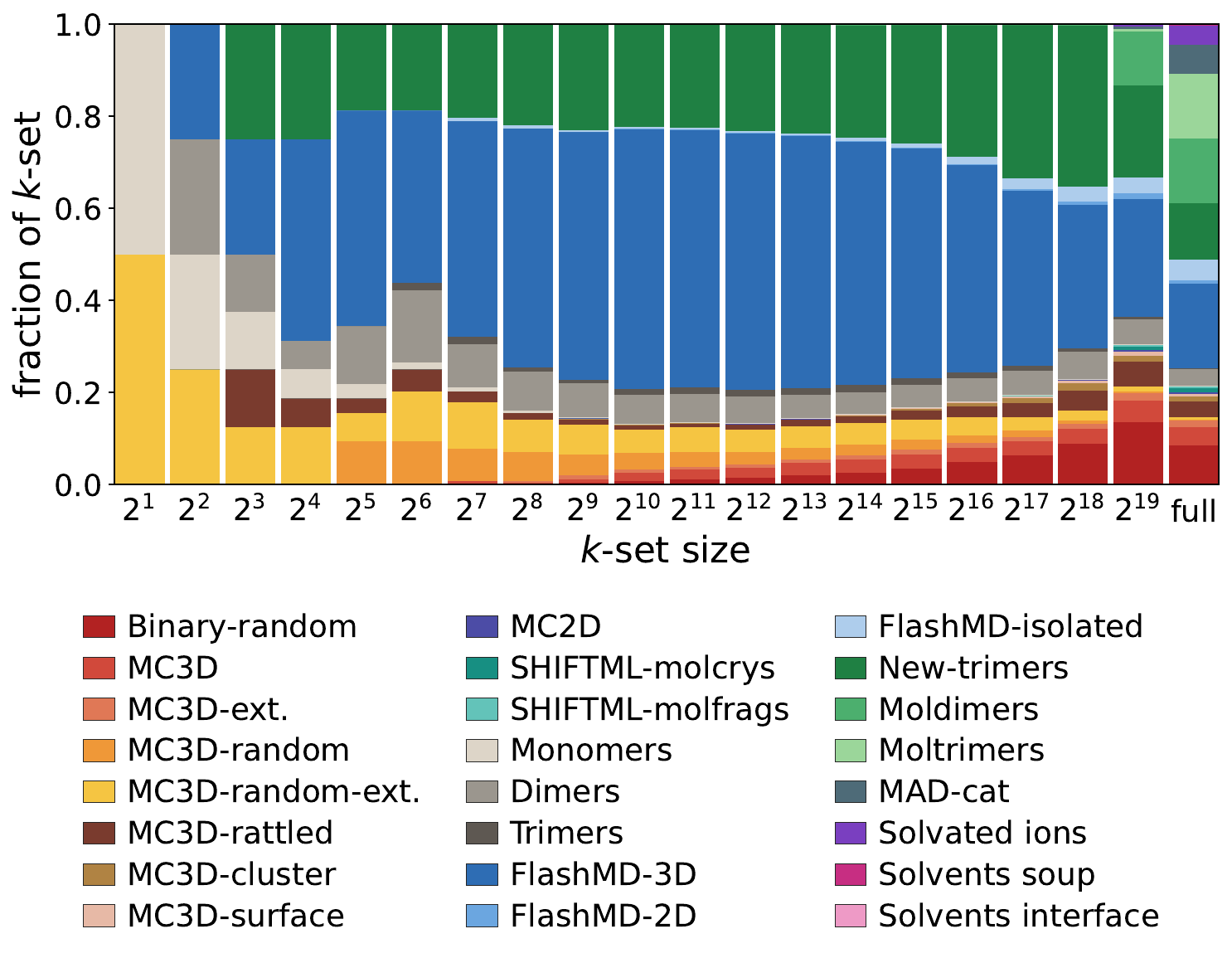}
    \caption{Decomposition of each MAD-NEXT selection by source subset. The bars correspond to nested power-of-two-sized coresets. The final bar is the full ordering of 856,012 MAD-NEXT structures}
    \label{fig:mad-coreset}
\end{figure}

\subsection{Coreset decomposition of MAD-NEXT}
\label{sec:mad-decomposition}

We seed the initial ordering by running FPS on MAD-NEXT starting from the hydrogen atom. 
MAD-NEXT is generated following the principle of maximizing chemical and structural diversity and using highly consistent \rrscan{} settings \cite{Furness2020, Malosso2026}. MAD-1.5 contributes 216,803 structures across 102 chemical elements and contains bulk crystals and their distorted subsets, low-dimensional and molecular systems, clusters, and isolated monomers, dimers and trimers. 
An additional 600,300 configurations include molecular dynamics snapshots of bulk and isolated systems, molecular clusters and fragments, and catalytic systems, and are complemented by 38,909 structures of solvated ions, bulk solvent, and solvent interfaces. 
Figure \ref{fig:mad-coreset} shows the decomposition obtained by running FPS  with \textit{ecumetric} on 856,012 MAD-NEXT structures into power-of-two nested sets, colored by source subset of each selection. The coarsest levels are dominated by very small configurations, monomers, dimers, and trimers, randomized crystals and FlashMD-3D snapshots \cite{bigi+25nips}, that is, configurations sampled along long-stride molecular dynamics trajectories of 3D periodic systems, which soon make up the largest part of the ordering. Up to $2^{12}$ selections, the majority of the picks contain no more than three atoms. The structures that the \textit{ecumetric} considers similar to each other, for example, molecular dimers and trimers, are represented by a few picks at the coarse levels and only saturate toward the end of the ordering.

\begin{figure}[htbp]
    \centering
    \includegraphics[width=\linewidth]{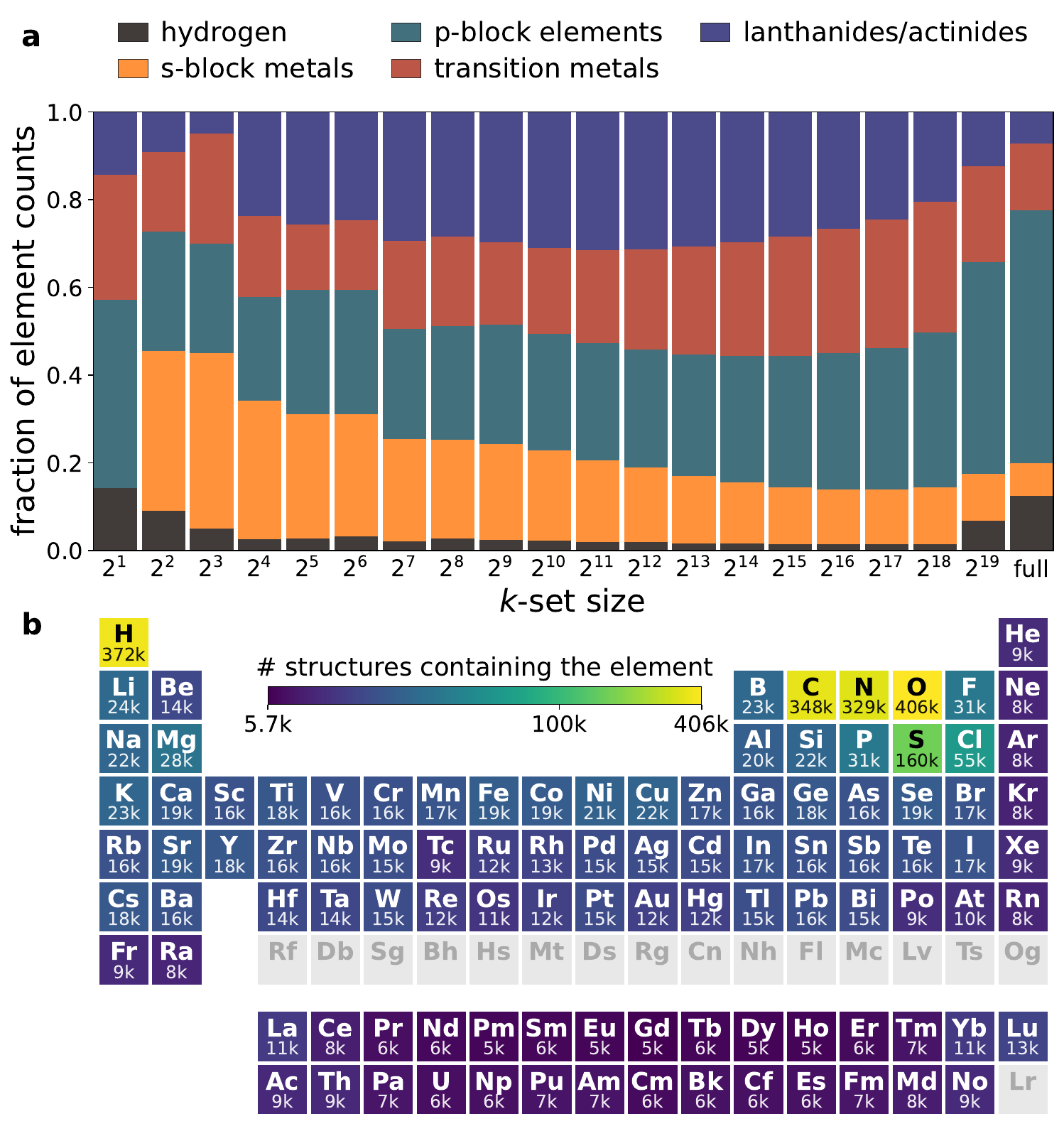}
    \caption{Elemental decomposition of MAD-NEXT coreset. a) Fraction of element counts by chemical family composition over selected power-of-two coresets. Each distinct element of the structure contributes one count. b) Number of structures
    containing each element over the full MAD-NEXT ordering. Each structure contributes once to every element present in its composition}
    \label{fig:mad-coreset-counts}
\end{figure}

The elemental composition changes as the hierarchy grows (Figure \ref{fig:mad-coreset-counts}a). The earliest layers are enriched in s-block metals, while lanthanides and actinides are picked a few times more often than their share in the full dataset, and the fraction of p-block elements gradually increases as coreset size becomes larger. Nevertheless, the first 480 picks already contain all 102 elements, but their fraction continues to change as more common carbon, oxygen, and nitrogen-containing configurations enter the coreset (Figure \ref{fig:mad-coreset-counts}b).

\begin{figure*}[htbp]
    \centering
    \includegraphics[width=\linewidth]{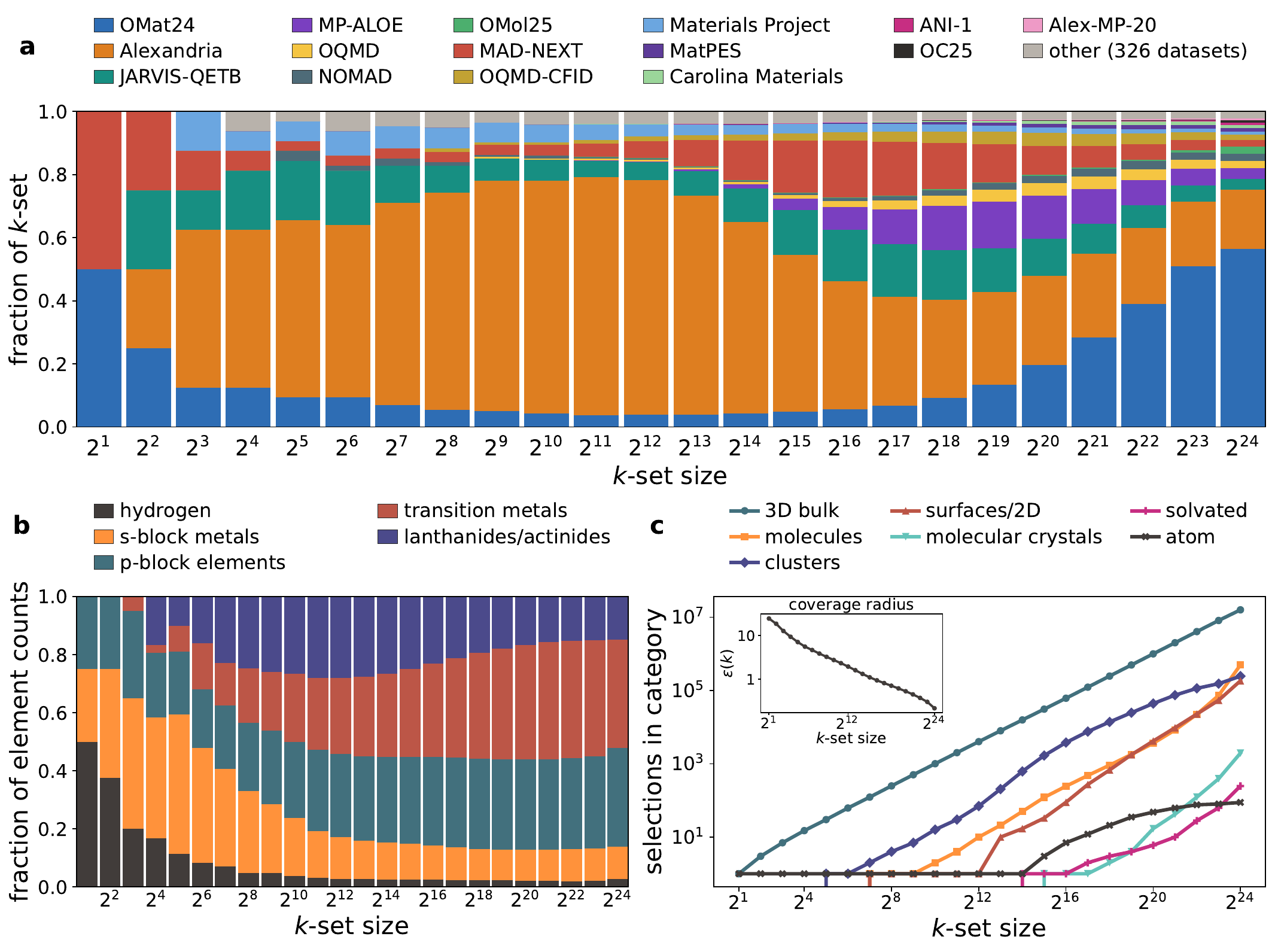}
    \caption{Decomposition of the $2^{24}$ library ordering. a) Fraction of each power-of-two coreset contributed by the fifteen largest dataset families; the remaining 326 datasets are grouped as other. b) Fraction of element counts by chemical family, each distinct element of a structure contributes one count. c) Cumulative number of selections in each structural category. The inset shows the coverage radius as a function of coreset size}
    \label{fig:coreset224-kset}
\end{figure*}

\begin{figure}[htbp]
    \centering
    \includegraphics[width=\linewidth]{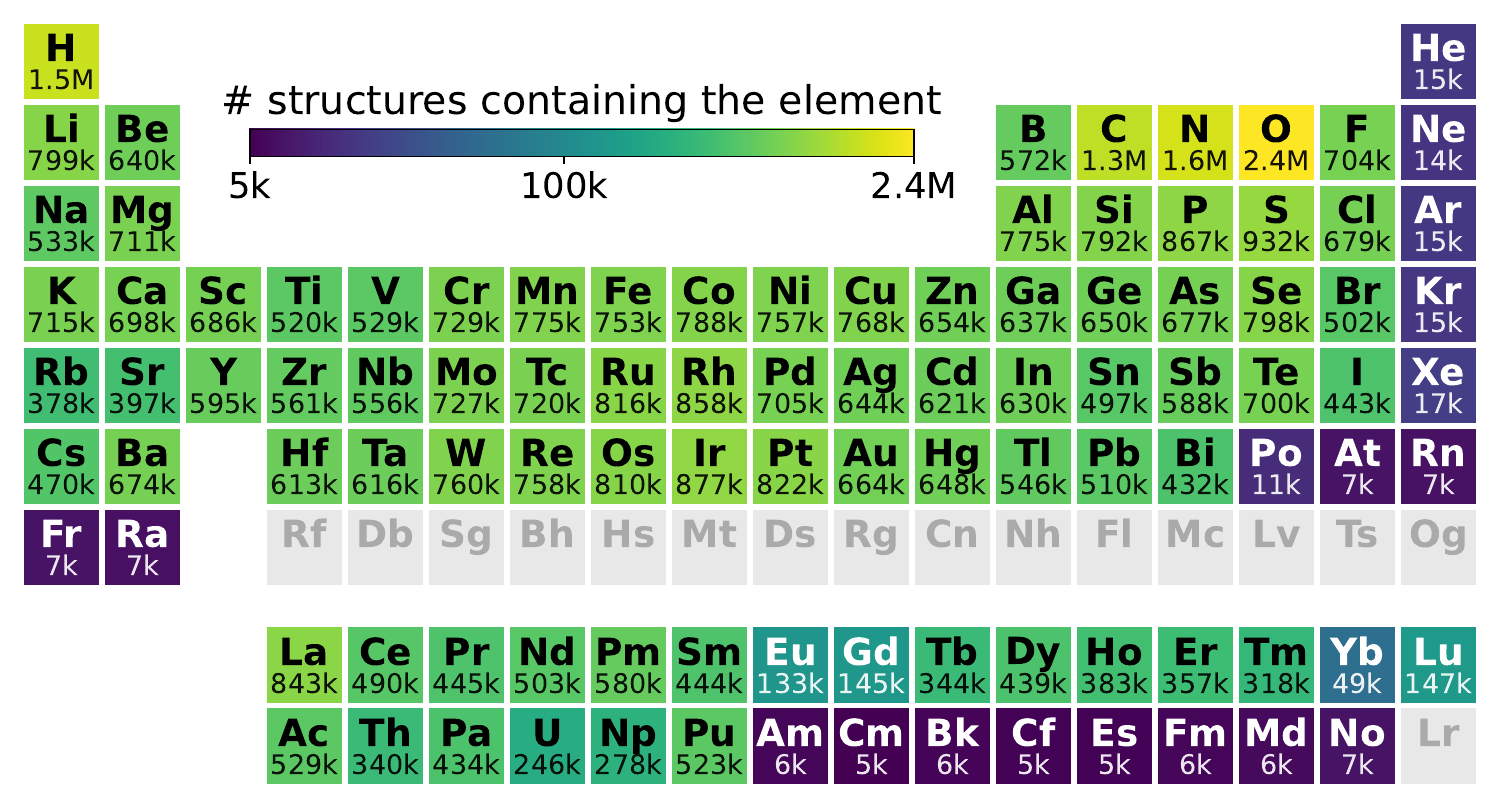}
    \caption{Number of structures containing each element over the full $2^{24}$ coreset; each structure contributes once to every element present in its composition}
    \label{fig:coreset224-elements}
\end{figure}

\subsection{Final coreset composition}

The accretion algorithm takes the 856,012 structures of the MAD-NEXT ordering as the anchor and extends it across the full candidate pool of 454 dataset families from MAD-COSMO.
Datasets that contain tens of millions of structures were first sub-selected with FPS up to 10M structures each. All pre-filtered datasets are then united into a single candidate pool of about 183 million structures, from which we proceed further.
Up to the size of the MAD-NEXT seed, accretion grows the hierarchy level by level, re-inserting as many of the original MAD-NEXT picks as the separation-coverage guarantee \eqref{eq:separation-coverage} permits, and filling the remaining slots by FPS over the pool. 
At this stage, about 10\% of the selected structures are retained from MAD-NEXT. This might seem a small fraction, but it is much larger than the share of MAD-NEXT in MAD-COSMO, which is about 2,000 times bigger.
Beyond that point, the selection simply continues as FPS over the whole pool, in which the remaining MAD-NEXT structures stay as candidates, up to a final depth of $2^{24} = 16,777,216$ structures. 
Beyond this level the selections begin to reflect the exhaustion of the capped FPS pools. Structures start to be picked not because they are needed to cover the space but because slots are freed once the large inorganic datasets pre-selected to 10M structures run out of candidates. Further refining of the coreset would increase the cost steeply, and would not make MAD-CORE much more useful, as further efforts to increase the resolution of a labeled dataset would probably have to be directed at specific positions of materials space.

\begin{figure*}[htbp]
    \centering
    \includegraphics[width=\linewidth]{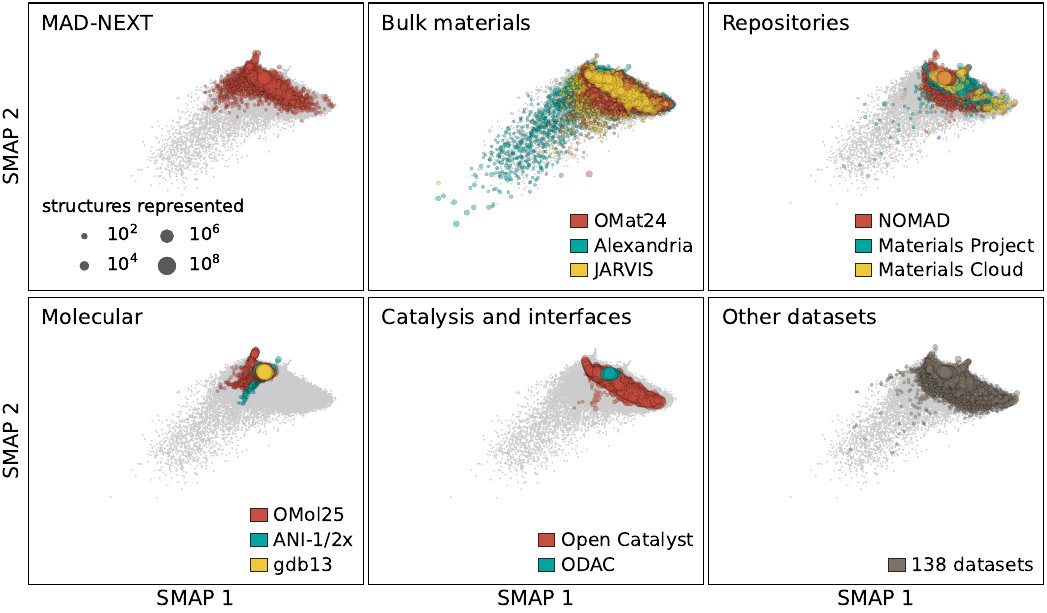}
    \caption{First two components of the eight-dimensional sketch-map projection of MAD-CORE. One million MAD-CORE structures sampled uniformly are shown in gray and one million sampled with probability proportional to $\sqrt{n}$ are shown in color. Marker size grows with the cell population $n$}
    \label{fig:smap}
\end{figure*}

Figure \ref{fig:coreset224-kset}a shows which datasets contribute to the selections at each power-of-two layer. Out of the 455 dataset families in the pool including MAD-NEXT, only 341 contribute a structure to the final coreset, the rest fall entirely within the coverage of earlier selections. Three-dimensional bulk materials make up 94\% of MAD-CORE, and molecules, clusters and surfaces almost all of the rest. OMat24 and Alexandria alone account for three quarters of the selections, followed by high-throughput crystal databases such as JARVIS-QETB, MP-ALOE and OQMD. The molecular corpus, from ANI, QM series, SPICE, OMol25 to the almost one billion structures of GDB-13, together makes up around three and a half percent of the final coreset, despite making up more than half of MAD-COSMO. The catalysis branch of datasets, OC20, OC22, OC25, makes up around one percent of all selections. Similarly to the MAD-NEXT decomposition, particular classes of configurations, such as molecular crystals, lie close to each other in the \textit{ecumetric} latent space and can be represented by a small number of points, in contrast to the far more extreme landscape of inorganic bulk materials, which the ordering continues to pick up for as long as it has budget, even though the molecular data contributes two thirds of our almost $1.7$ billion collection.
340,555 MAD-NEXT structures end up in the final ordering despite making up less than 1\% of the pool and less than 0.1\% of MAD-COSMO---a demonstration of the high degree of structural and chemical diversity.
The subsets with no well-covering counterparts in the rest of the pool, random binary compounds, rattled crystals, molecular trimers and FlashMD snapshots, keep most of their picks. MAD-NEXT solvated systems and equilibrium MC3D crystals are largely replaced by much larger dedicated sources that cover those regions more broadly.

Beyond the first few thousand selections the relative contributions of the chemical families remain nearly constant (Figure \ref{fig:coreset224-kset}b). Transition metals and p-block elements take almost one third of the element counts each and lanthanides and actinides are picked early, but their share shrinks as the coreset grows. The selection starts with the same isolated hydrogen atom that seeds the MAD-NEXT decomposition, and bulk crystals, clusters, molecules and surfaces all appear within roughly the first hundred picks. Even the rarest categories, solvated systems and molecular crystals, are represented within the first $2^{15}$ selections (Figure \ref{fig:coreset224-kset}c). All 102 elements of the collection are covered within the first 2,158 selections and the element frequency in the full coreset (Figure \ref{fig:coreset224-elements}) ranges from the millions for common organic elements down to a few thousand for the heaviest actinides and the short-lived radioactive elements.

Only 26 out of 341 datasets contribute more than ten thousand structures, and 235 place fewer than 100 structures, which means that much of the breadth of the coreset consists of small contributors. For example, individual specialized Materials Cloud and NOMAD uploads together add over 400,000 structures. Nevertheless, the library organizes what the community has already computed and therefore cannot cover the regions that no dataset has sampled. Approaches that generate maximally diverse configurations from scratch, such as the SMAX dataset \cite{Bochkarev2026}, could help fill these gaps. If folded into the pool, such configurations would end up early in the ordering if they occupy regions not covered by the existing datasets in the library.

\subsection{Voronoi tessellation}
\label{sec:voronoi}

The coreset represents the library through its most diverse members. To relate the rest of the collection to them, we assign each structure to its nearest coreset point at each level of the hierarchy, i.e. we create a Voronoi tessellation, as introduced in Section \ref{sec:accrete}. We additionally FPS-order the points within the cells of the finest level ($2^{24}$) of the coreset.

The separation-coverage guarantee \eqref{eq:separation-coverage} holds for the pool of structures the FPS ran on, and the pool itself is a capped subsample of the whole collection, so one might wonder how far the remaining structures actually are from the coreset. In practice, they are very close:
taking for reference the coverage radius reached on the pool, $\epsilon(2^{24}) = 0.218$, and considering only the structures that passed the ZBL sanity checks, the largest distance between a structure of the full library and its nearest MAD-CORE point (the directed Hausdorff distance) is 0.4, $99.17\%$ of the full library lies within $\epsilon(2^{24})$, and $99.93\%$ within the second finest coreset coverage level $\epsilon(2^{23}) =0.304$.
In other terms, the coreset covers the structures it has never seen almost as tightly as the pool it was built from, justifying our strategy of selecting at most 10 million structures from each of the largest datasets. At the deepest level, half of the collection falls into just 117 of the almost seventeen million cells, and the single largest cell gathers 115 million configurations around one 26-atom organic molecule from GDB-13.

This Voronoi tessellation can also be helpful to mitigate the limitations of using FPS training set selection \cite{bart+17sa, Trestman2025}. Sampling the coreset points weighted by their cell populations interpolates between the two regimes of uniform coverage and density based selection \cite{ceri+13jctc}. These populations show how often similar configurations occur in the source datasets, rather than their probability under particular physical conditions.
Considering this partitioning useful for further large-scale campaigns of dataset creation, we share the tessellation together with all other coreset data for all points used in this study (Section \ref{sec:data-availability}).

\subsection{Data visualization}\label{sub:sketchmap}

Adopting the approach demonstrated in the MAD-1 paper \cite{mazi+25sd}, we use sketch-map \cite{ceri+11pnas} nonlinear dimensionality reduction, which matches pairwise distances around a chosen scale between the full and reduced dimensions. We fit the map on the first $2^{20}$ coreset points to project the 512-dimensional \textit{ecumetric} features to eight dimensions, and weight each point in the stress by a power of the population of its Voronoi cell, so that the map reflects how the $1.7$ billion structures are arranged in the reduced \textit{ecumetric} space. Building on Ref.~\cite{mazi+25sd}, we train a projector that extends the map to unseen structures and provide it as a \texttt{metatomic} model \cite{bigi+26jcp}. The model computes the \textit{ecumetric} features of an input structure and projects them onto the reduced feature space fitted on MAD-CORE. The model can be used with the \texttt{chemiscope.explore} module of Chemiscope \cite{Chorna2026Chemiscope} to project new structures and assess whether they are already well represented or occupy regions of the feature space that remain sparsely sampled. The parameters of the map and details of the projector are given in SI \ref{app:sketchmap}.
The resulting 2D map (Figure \ref{fig:smap}) can be used to extract some insights about the ground covered by different datasets, even though one should keep in mind that this is a highly coarse-grained representation that does not fully reflect the similarity between configurations.
MAD-NEXT covers a respectable portion of space (especially considering its size), but there is a large region, consisting mostly of bulk materials, that extends beyond it. This tail is best covered by Alexandria, which is the most diverse among the large datasets of inorganic materials. 
Molecular datasets cover a small region, with the most striking observation being that GDB-13, with almost 1B structures, is concentrated in a tiny portion of chemical space.
This observation reflects the more limited compositional space (GDB-13 contains exclusively H, C, N, O, S, Cl) and is consistent with only a few structures being retained in MAD-CORE. It also suggests that great care should be taken when using OMol25-trained models to make predictions for bulk structures, as they miss a lot of the structural and chemical diversity of inorganic datasets. 
Finally, the small remaining datasets have a lot to bring, contributing both in terms of fine-grained coverage of the high-density part of the bulk materials datasets, and in bringing a few outliers that reach into the sparsely-populated part of the map. 

\begin{figure}[bp]
    \centering
\includegraphics[width=\linewidth]{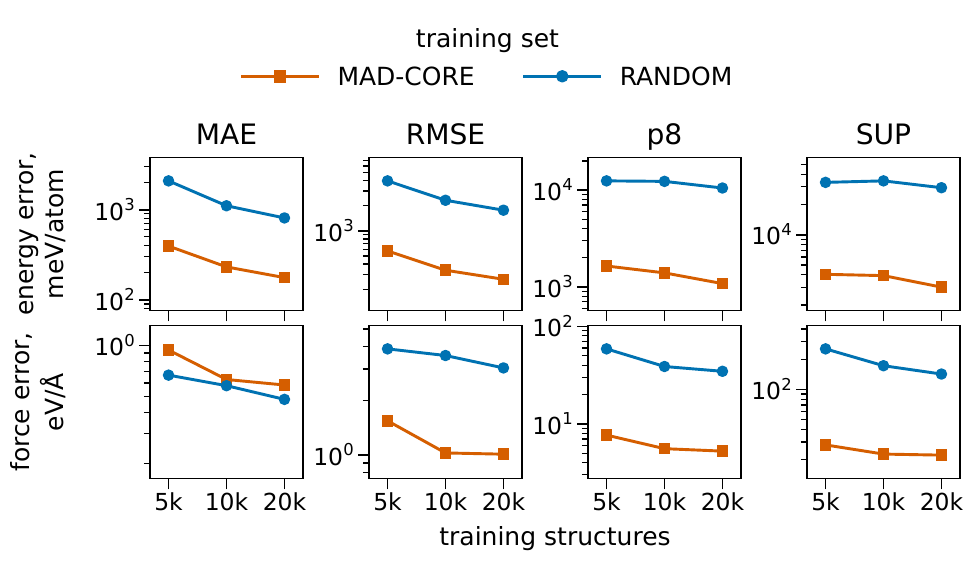}
    \caption{Learning curves of PET models trained from scratch on progressively larger MAD-CORE and random MAD-COSMO sets, evaluated on a test set of 2k random and 2k held-out MAD-CORE structures. Columns show MAE, RMSE, and the tail-sensitive eighth-order mean error ($p_8$) and maximum absolute error (SUP) for energy and forces}
    \label{fig:learning_curves}
\end{figure}

\section{MAD-CORE as a compact training set}
\label{sec:training}

Generating large datasets of high-quality quantum properties requires enormous computational investments. To demonstrate how MAD-CORE can help focus these efforts, we recomputed up to 20 thousand structures with the PBE density functional~\cite{perd+96prl}, selected from either the top levels of the MAD-CORE ordering or randomly from the entirety of MAD-COSMO, using the all-electron FHI-aims code \cite{Abbott2026, blum+13cpc}.
We exclude structures with 300 or more atoms or with a maximum atomic force magnitude above 100 eV/\AA{}.
We then train on the two selections, and on subsets of 5 and 10 thousand structures, machine-learning interatomic potentials based on the PET architecture~\cite{pozd-ceri23nips}. 
We include the 102 isolated atom energies in each training set, and use the same training protocol and number of optimizer updates for each model. 
We use as test set the combination of 2000 random structures from MAD-COSMO and 2000 MAD-CORE structures.  When comparing results, one should note that even though the random and hierarchical training sets contain the same number of structures, random structures are typically larger, so that on average the MAD-CORE-derived training set contains approximately six times fewer force labels. Detailed training and electronic-structure protocols are given in SI \ref{app:train_mad-core}.

\begin{figure}[tbp]
    \centering
    \includegraphics[width=\linewidth]{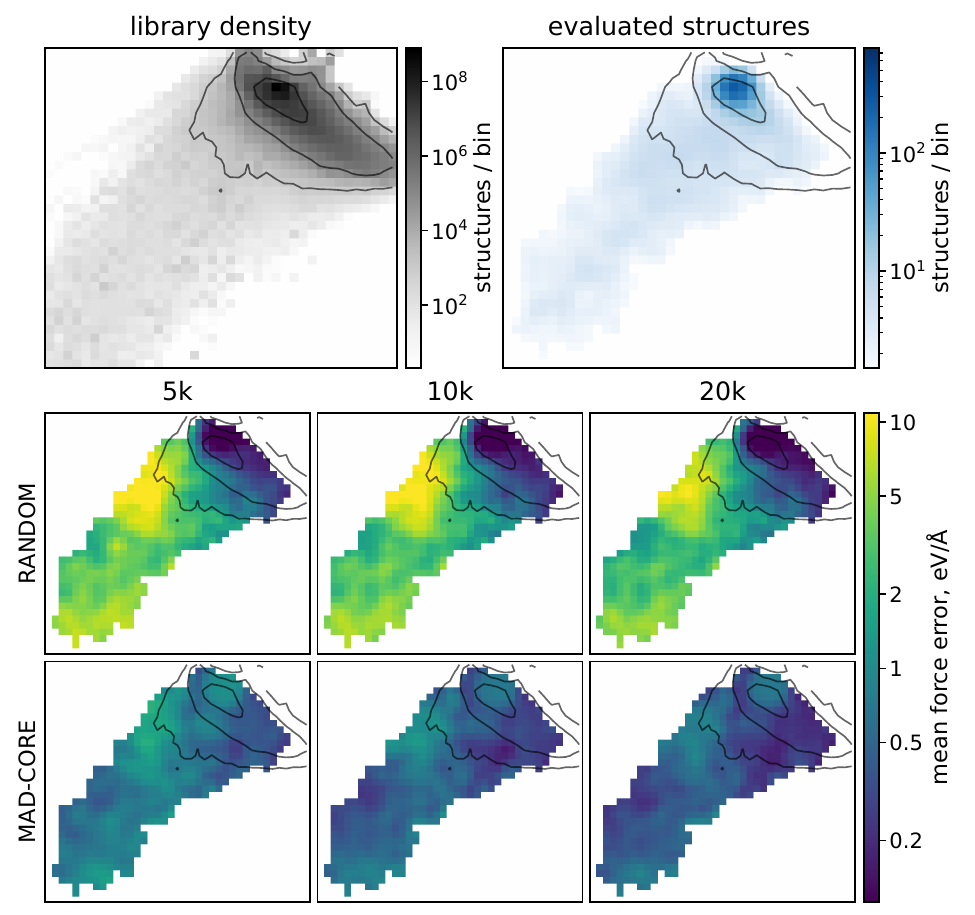}
    \caption{The sketch-map projection of the common test set colored by the force errors. The top panel shows MAD-COSMO population density and the density of the test set. Below, the panels show smoothed local mean force errors of PET models trained on progressively larger MAD-CORE or random training sets}
    \label{fig:error_map}
\end{figure}

The learning curves show that both choices of sample selection lead to errors that decay with dataset size, but the relative performance depends on the choice of metric.
The advantage of MAD-CORE is clearest when using metrics that emphasize extreme cases, such as the supremum norm, or a high $p$-norm. 
For outlier-insensitive metrics, and especially for the force MAE, for which the randomly-selected set contains six times more data, random training is competitive. 
Learning curves evaluated separately on the random and hierarchical sets are given in the SI~\ref{app:train_mad-core}, and show that (unsurprisingly) each model tends to perform better on the test set that has a comparable statistical makeup.
The effect can be seen most clearly by resolving the errors based on the position of each sample on the 2D map of the dataset (Figure \ref{fig:error_map}). 
Random training yields low errors in the densely populated part of chemical space, but enormous errors at the periphery. Conversely, MAD-CORE training yields errors that are much more uniform across the full test set, and improve across the board with increasing amounts of data.
We argue that models striving for universal applicability, and the datasets supporting them, should prioritize robustness to outliers over the MAE metrics that are usually chosen for benchmarks.
Mixing random and MAD-CORE sampling, or using scaled-Voronoi-density sampling with a low exponent, provide viable strategies to interpolate between the two limiting cases.

These tests, training models from scratch on minuscule databases of 20k structures, are meant to compare different sampling strategies with the least amount of confounding factors, but it is clear that the resulting accuracy is insufficient to use the models confidently. 
There are however several practical approaches to reduce errors dramatically. 
First, one can use a pre-trained model as the starting point rather than training from scratch. 
On the same test set, fine-tuning of PET-OMat-S \cite{Bigi2026PushingTheLimits} on 20k MAD-CORE structures reduces energy and force MAEs to 90 meV/atom and 191 meV/\AA{}, compared with 177 meV/atom and 583 meV/\AA{} when training from scratch. 
In the most relevant scenario, that of transfer learning between levels of theory, one can additionally learn the difference between the two sets of targets, a well established baselining strategy sometimes referred to as $\Delta$-learning \cite{rama+15jctc} that reduces the variance and facilitates learning from small amounts of data. 
To demonstrate the application of MAD-CORE in this scenario, we use the top 18k MAD-CORE structures, computed with both the r$^2$SCAN functional~\cite{Furness2020} and the local density approximation (LDA)~\cite{kohn-sham65pr} (results for a PBE baseline are given in the SI).
Directly fine-tuning PET-OMat-S on the r$^2$SCAN data yields energy and force MAEs of 143 meV/atom and 382 meV/\AA{}  on the MAD-CORE holdout test set, while the difference between the two functionals has errors of just 29 meV/atom and 62 meV/\AA{}---respectable performance over such diverse test structures.
On the random MAD-COSMO test set, the corresponding errors decrease from 62 meV/atom and 266 meV/\AA{} to 20 meV/atom and 41 meV/\AA{}. 
The learned $\Delta$ corrections can be subtracted from an accurate r$^2$SCAN model, such as PET-MAD-1.6-M~\cite{Malosso2026}, to run stable simulations with an approximate LDA potential (see SI~\ref{app:delta-md}). These simulations reproduce the (physically incorrect) structural properties of LDA water, and the corrected potential shows small errors against DFT calculations for a few snapshots selected along the trajectories.
Training a correction to a \emph{lower} level of theory is clearly only a computationally affordable proof of principle of the approach, but it provides the evidence to motivate further efforts, both by ourselves and the broader community, to generate data to train universal models that can deliver hybrid DFT accuracy across the full chemical space covered by MAD-COSMO.

\section{Conclusions}

We consolidate more than one thousand individually published, publicly available atomistic data records, containing a total of about 1.7 billion structures, into a single collection, MAD-COSMO. 
We organize it into a nested coreset hierarchy, MAD-CORE, consisting of the most diverse $2^{24}$ structures. At each level of the hierarchy the first $2^{k}$ structures form a coreset that covers the structures it was selected from uniformly at its own resolution. The construction relies on FPS in the space of the \textit{ecumetric}, a structural metric trained on the features of several uMLIPs taken on the MAD-NEXT dataset, independently built to achieve high structural and chemical diversity. 
The ordering is seeded with an FPS run over MAD-NEXT and extended by an accretion algorithm, designed to incorporate new data with minimal changes to the coarser levels of the hierarchy, followed by plain FPS at the finest levels. A Voronoi tessellation assigns every structure of the library to its nearest coreset representative at all levels of the hierarchy. 
More than $99\%$ of the collection lies within the coreset coverage radius even though the selection never saw most of it. The resulting coverage remains conditional on the representation, curation choices, and available structures but provides a common basis for organizing and extending atomistic datasets.

In practice, MAD-CORE can be used as subsets for recalculation at higher levels of theory with consistent electronic-structure settings, benchmarking, and training-set construction. 
The latter can be balanced between diversity and match to the underlying MAD-COSMO sampling density by a weighted selection based on the populations of the Voronoi cells generated by MAD-CORE. 
A trained sketch-map projector places any new structure within the reduced dimensions of the \textit{ecumetric} space to show whether it is already well represented. For every MAD-CORE structure we also distribute its \textit{ecumetric} features and the SPACE-predicted energies and non-conservative forces, which can be used to pre-train models on a large amount of synthetic data.

Training on MAD-CORE subsets of increasing size leads to ML potentials that exhibit near-uniform accuracy across the full space covered by MAD-COSMO, with outlier-sensitive metrics being up to 10 times better than for models trained on randomly-selected structures. This comes however at the expense of the accuracy for randomly-selected test sets, which is to be expected given the wildly inhomogeneous sampling density in MAD-COSMO.
We demonstrate that a comparatively small number of energy and force calculations at the coarser coreset levels make it possible to train accurate and stable corrections to an underlying universal force field trained on a different level of electronic-structure theory. 
This proof of principle motivates a collective effort to compute high-accuracy labels for MAD-CORE, as well as to further extend it while avoiding redundant calculations, going from a passive implementation of FAIR principles that makes data available to a more active strategy in which open data records are used to coordinate and drive community actions.

\section*{Data and code availability}
\label{sec:data-availability}

\begin{table*}[tbp]
\caption{Overview of the MAD-CORE and MAD-COSMO data records. Energies are in eV, forces in eV/\AA{}, coordinates and cells in \AA{}, stresses in eV/\AA{}$^3$. MAD-CORE structures follow the coreset ordering, while MAD-COSMO structures are grouped by their corresponding MAD-CORE Voronoi cells. The structure and feature files are partitioned using the same intervals of MAD-CORE indices: $[0,2^{18})$, $[2^{18},2^{19})$, \ldots, $[2^{23},2^{24})$.
}
\label{tab:records}
\begin{ruledtabular}
\begin{tabular}{p{0.13\linewidth}p{0.19\linewidth}p{0.13\linewidth}p{0.47\linewidth}}
Component & Entries & Format & Contents \\
\colrule
\multicolumn{4}{l}{
\textbf{MAD-CORE record}~\cite{madcore_matcloud}} \\
\raggedright Structures
& \raggedright $2^{24}=16{,}777{,}216$ structures
& \raggedright Extended XYZ, gzip & \raggedright Atomic numbers, positions, cell; \textit{ecumetric} SPACE energy and non-conservative forces; configuration ID, source dataset ID and group, ID in the source dataset; global FPS rank, smallest level $k$ containing the structure, squared FPS distance $\epsilon^2$, Voronoi-cell population in MAD-COSMO; ZBL filtering diagnostics
\tabularnewline

\raggedright Features
& \raggedright One 512-dimensional vector per structure
& \raggedright HDF5, float32
& \raggedright \textit{Ecumetric} features
\tabularnewline

\raggedright DFT labels
& \raggedright 79,241 LDA, 77,383 PBE, 21,854 r$^2$SCAN structures
& \raggedright Extended XYZ, gzip
& \raggedright FHI-aims energies, free energies, energies relative to the isolated atoms, forces, and stresses (where available). LDA/PBE: converged calculations from four selections: the first 25,000 MAD-CORE structures; structures from 25,000 Voronoi cells drawn with probability proportional to $n^{1/4}$, where $n$ is the cell size; 25,000 random MAD-COSMO structures; 6,000 random MAD-COSMO test structures. r$^2$SCAN: 18,000 training, 1,000 validation and 1,881 test
structures from the top of MAD-CORE, and 973 random MAD-COSMO test structures, split assignments stored in the files
\tabularnewline

\raggedright Isolated atoms
& \raggedright 102 elements per functional
& \raggedright XYZ
& \raggedright LDA, PBE and r$^2$SCAN energies of isolated atoms
\tabularnewline

\raggedright DFT settings
& \raggedright One template per functional
& \raggedright \texttt{control.in}, tar.gz
& \raggedright FHI-aims input templates for LDA, PBE and
r$^2$SCAN
\tabularnewline

\raggedright Sketch-map & \raggedright First $2^{20}$ structures & \raggedright Chemiscope JSON, gzip & \raggedright Eight-dimensional sketch-map projection of the \textit{ecumetric} features
\tabularnewline

\raggedright Dataset metadata & \raggedright One row per source dataset & \raggedright CSV & \raggedright Title, upstream URL, source publication, license
\tabularnewline

\colrule
\multicolumn{4}{l}{\textbf{MAD-COSMO record}~\cite{madcosmo_matcloud}} \\
\raggedright Structures & \raggedright ${\approx}1.7\times10^{9}$ structures, grouped in the Voronoi cells of the $2^{24}$ MAD-CORE structures & \raggedright HDF5 & \raggedright Atomic numbers, positions, cell; \textit{ecumetric} SPACE energy and non-conservative forces; configuration ID, source dataset ID and group, ID in the source dataset; FPS rank within the cell and squared FPS distance $\epsilon^2$; nearest MAD-CORE structure among the first $2^k$, and its distance, for $k=0,\ldots,24$; ZBL filtering diagnostics
\tabularnewline

\raggedright Features & \raggedright One 512-dimensional vector per structure & \raggedright HDF5, float32 & \raggedright As in the MAD-CORE record \tabularnewline
\raggedright Dataset metadata & \raggedright One row per source dataset record & \raggedright CSV & \raggedright As in the MAD-CORE record, plus the number of structures and the retrieval channel of each dataset
\tabularnewline
\end{tabular}
\end{ruledtabular}
\end{table*}

We release MAD-CORE as a record \cite{madcore_matcloud} at Materials Cloud Archive \cite{tarl+20sd}. It is shared as compressed extended-XYZ files organized telescopically over power-of-two levels. It holds the atomic numbers, positions, and cell, the provenance of each structure, including its identifier, the source dataset and the identifier needed to retrieve the original source entry, energy and non-conservative forces predicted by the \textit{ecumetric} SPACE model, and ZBL filtering diagnostics. We also share the structure-level \textit{ecumetric} features as HDF5 files. The LDA, PBE and r$^2$SCAN calculations performed for this work are also included in the record. Reference calculations from the original source datasets are not redistributed but can be retrieved using the provenance fields. The metadata records the licenses and provenance of each source dataset.
The MAD-COSMO data, including the Voronoi tessellation with within-cell ordering, is included in a separate Materials Cloud record. 
MAD-NEXT labels are not released in this record but will be available as part of MAD-2, subject to curation.
The content and format of both records are summarized in Table~\ref{tab:records}.

The \textit{ecumetric} model and the sketch-map projector are available at \url{https://huggingface.co/lab-cosmo/ecumetric}. The sketch-map and CPU FPS are part of the API of \texttt{scikit-matter} \cite{gosc+23ore}. CUDA and Python implementations of multi-GPU FPS and \textit{accrete} are available at \url{https://github.com/sofiia-chorna/coresets}. The SPACE model architecture is part of \texttt{metatrain} \cite{bigi+26jcp}. The scripts for the analysis and figures are archived at Zenodo \cite{madcore_code}.

\begin{acknowledgments}
We thank the authors of MAD-2 for providing early access to a preliminary version of the dataset~\cite{mad-next}, and the authors of all datasets whose open licensing made this work possible. We acknowledge computing resources provided by the Swiss National Supercomputing Centre under project aa002 and by SCITAS at EPFL.
SC and MC were supported by the NCCR MARVEL a National Centre of Competence in Research, funded by the Swiss National Science Foundation (grant number 205602), and by the SNSF project n. 200020\_214879.
JS acknowledges support from the Swiss National Science Foundation (SNSF) under Ambizione grant number 233444. 
CM and MC acknowledge support by the NCCR Separations, a National Centre of Competence in Research funded by the Swiss National Science Foundation (grant number 229280).

\end{acknowledgments}

\section*{Author contributions}
S.C. developed the coreset construction, co-curated the collection, trained the \textit{ecumetric} and sketch-map projector, performed the validation experiments, prepared the data release, and wrote the initial draft. C.M. and S.C. performed the electronic-structure calculations. C.M. contributed to the formulation of the hierarchical selection scheme. J.S. developed the \textit{ecumetric} approach, trained its initial version, collected and curated datasets, and supervised the work. M.C. designed and guided the project and provided theoretical support. All authors contributed to the writing and revisions.
The authors acknowledge Claude (Anthropic) and ChatGPT (OpenAI) for assistance with code development and manuscript preparation, and take full responsibility for the resulting implementation and text.

%

%
%
%
%

%
%
%

\onecolumngrid\clearpage
\includepdf[fitpaper, pages={{},-}, pagecommand={\thispagestyle{empty}\clearpage}]{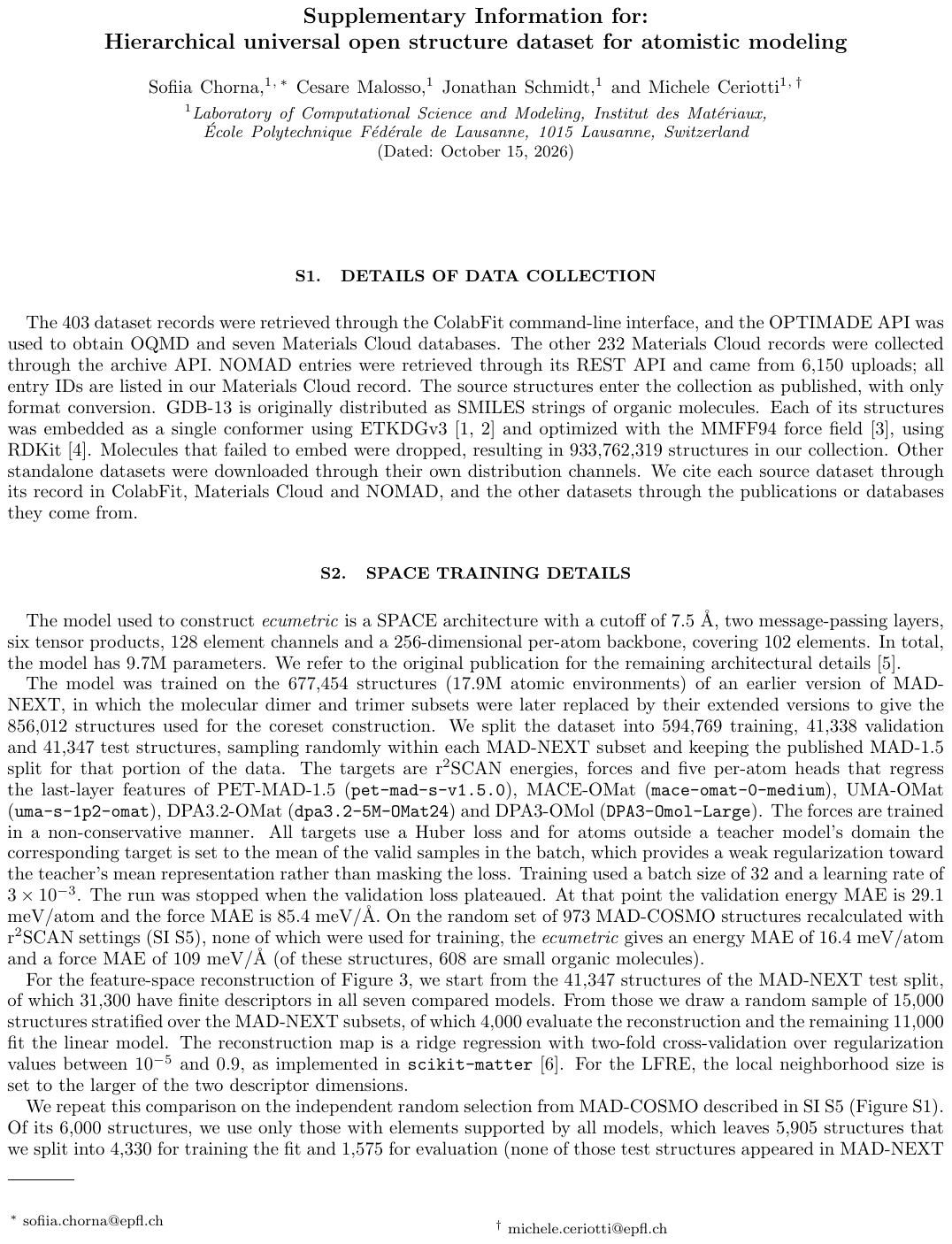}

\end{document}